\documentclass{aa}
\usepackage[varg]{txfonts}

\newcommand{\bc}{\begin{center}}
\newcommand{\ec}{\end{center}}
\newcommand{\be}{\begin{equation}}
\newcommand{\ee}{\end{equation}}
\newcommand{\bfig}{\begin{figure}}
\newcommand{\efig}{\end{figure}}
\newcommand{\m}{\mbox}

\newcommand{\oz}{O$_3$}
\newcommand{\om}{O$_2$}
\newcommand{\od}{O($^1$D)}

\makeatletter
\newcommand*{\linktocite}[2]{%
  \hyper@natlinkstart{#1}#2\hyper@natlinkend}
\makeatother

\makeatletter
\renewcommand*\aa@pageof{, page \thepage{} of \pageref*{LastPage}}
\makeatother

\begin{document}

\title{Is ozone  a reliable proxy for molecular oxygen?} 
\subtitle{I. The \om-\oz\ relationship for Earth-like atmospheres}

\author{Thea Kozakis \inst{1}
  \and Jo\~ao M. Mendon\c{c}a \inst{1} 
  \and Lars A. Buchhave \inst{1}} 
  
\institute{National Space Institute, Technical University of Denmark, Elektrovej, DK-2800 Kgs. Lyngby, Denmark}

\date{}							

\abstract
{Molecular oxygen (\om) paired with a reducing gas is regarded as a promising biosignature pair for the atmospheric characterization of terrestrial exoplanets.  In circumstances when \om\ may not be detectable in a planetary atmosphere (e.g., at mid-IR wavelengths) it has been suggested that ozone (\oz), the photochemical product of \om, could be used as a proxy to infer the presence of \om.  However, \oz\ production has a nonlinear dependence on \om\ and is strongly influenced by the UV spectrum of the host star.  To evaluate the reliability of \oz\ as a proxy for \om, we used \texttt{Atmos}, a 1D coupled climate and photochemistry code, to study the \om-\oz\ relationship for ``Earth-like'' habitable zone planets around a variety of stellar hosts (G0V-M5V) and \om\ abundances. Overall, we found that the \om-\oz\ relationship differed significantly with stellar hosts and resulted in different trends for hotter stars (G0V-K2V) versus cooler stars (K5V-M5V).  Planets orbiting hotter host stars counter-intuitively experience an increase in \oz\ when \om\ levels are initially decreased from 100\% Earth's present atmospheric level (PAL), with a maximum \oz\ abundance occurring at 25-55\% PAL \om. As  \om\ abundance initially decreases, larger amounts of UV photons capable of \om\ photolysis reach the lower (denser) regions of the atmosphere where \oz\ production is more efficient, thus resulting in these increased \oz\ levels. This effect does not occur for cooler host stars (K5V-M5V), since the weaker incident UV flux does not allow \oz\ formation to occur at dense enough regions of the atmosphere where the faster \oz\ production can outweigh a smaller source of \om\ from which to create \oz.  Thus, planets experiencing higher amounts of incident UV possessed larger stratospheric temperature inversions, leading to shallower \oz\ features in planetary emission spectra.  Overall it will be extremely difficult (or impossible) to infer precise \om\ levels from an \oz\ measurement, however, with information about the UV spectrum of the host star and context clues, \oz\ will provide valuable information about potential surface habitability of an exoplanet.
}


\titlerunning{\oz\ as a Proxy for Molecular Oxygen I}
\authorrunning{Kozakis, Mendon\c{c}a, \& Buchhave et al.}

\maketitle

\section{Introduction}

In the search for life in the Universe, molecular oxygen (\om) is commonly recognized as a promising atmospheric biosignature gas.  However, while \om\ is largely created by biological sources on Earth, it can also be produced abiotically in a variety of settings, and thus alone would not constitute a guarantee of life (e.g., \citealt{hu12,word14,doma14,tian14,luge15,gao15,harm15}).  Instead of being a standalone biosignature, \om\ as a biosignature will be most powerful when detected simultaneously with a reducing gas as a ``disequilibrium biosignature pair'' (e.g.,  \citealt{love65,lede65,lipp67}), and when evidence of abiotic \om\ production scenarios can be ruled out (see \citealt{mead17,mead18} for a review).

In scenarios where \om\ is not directly detectable, it has been suggested that its photochemical product ozone (\oz) could be used as a proxy for \om\ (e.g., \citealt{lege93,desm02,segu03,lege11,mead18}).  Using \oz\ as a proxy for \om\ would be extremely useful in two particular scenarios: 1) at wavelengths where \om\ features are not present (i.e., mid-infrared wavelengths), and 2) when \om\ is present in small amounts (as it was for a significant fraction of Earth's geological history).

The mid-infrared wavelength region (MIR; 3-20 $\mu$m) provides an excellent opportunity for the search for life, as it contains features for multiple biosignature gases, as well for gaseous species that could provide evidence for or against biological \om\ production \citep{desm02,schw18,quan21}.  Furthermore, thermal emission observations are less impacted by clouds (e.g., \citealt{kitz11}), and could also allow measurements of a planet's surface temperature \citep{desm02}. 
The collisionally-induced absorption \om\ feature at 6.4 $\mu$m is the only MIR feature that allows for the direct detection of \om, although it would be extremely difficult to use for abundances of \om\ consistent with biological production \citep{fauc20}.  It will, however, be useful for identifying high-\om\ desiccated atmospheres, a possible mechanism for abiotic \om\ production \citep{luge15,tian15}.
Inferring the presence of biologically produced \om\ will be restricted to indirect detections via the 9.7 $\mu$m \oz\ feature in the MIR.

In addition, although \om\ has existed in appreciable amounts on Earth for a significant part of its history, it has only existed in large amounts for a relatively short period of time, posing a fundamental drawback to \om\ as a biosignature \citep{mead18}.  Molecular oxygen was first created produced biologically $\sim$2.7 Ga (billion years ago) by oxygenic photosynthesis via cyanobacteria,  although it did not build up to appreciable amounts in Earth's atmosphere until the Great Oxidation Event (GOE) $\sim$2.45 Ga (see e.g., \citealt{catl17} for a review).  Although the Phanerozoic era (541 Ma - present day) saw the widespread colonization of land plants and \om\ levels comparable to our present atmospheric level (PAL), during the Proterozoic era (2.5 Ga - 541 Ma) it is expected \om\ levels could have been significantly lower  \citep{catl17,lent17,dahl20}.  As a result, it is likely that \om\ only would have been detectable on Earth for the last $\sim$0.5 Gyr.  However, since \oz\ is a logarithmic tracer of \om, it is possible that \oz\ could be capable of revealing small, undetectable amounts of \om\  (e.g., \citealt{kast85,lege93,desm02,segu03,lege11}). Additionally, a detection of \oz\ could provide information about UV shielding, and whether surface life is adequately protected from high-energy UV capable of DNA damage.

Some studies have suggested or already adopted O$_3$ as a substitute for O$_2$ (e.g.,\ \citealt{segu03}, \linktocite{KalteneggerMacDonald2020}{Kaltenegger \& MacDonald et al.} \citeyear{KalteneggerMacDonald2020},\ \citealt{lin21}), and others have noted a potentially powerful ``triple biosignature'' in planetary emission with CO$_2$, H$_2$O, and O$_3$, where O$_2$ spectral features are absent \citep{sels02}. O$_3$ is also expected to build up in the stratospheres of planets, allowing characterization via transmission spectroscopy (e.g., \citealt{betr13,betr14,misr14,mead18}).

 However, it is uncertain how reliably a measurement of \oz\ could allow us to infer the amount of \om.  Ozone is known to have a nonlinear relationship with \om, as well as a strong dependence on the UV spectrum of the host star \citep{ratn72,kast80,kast85,segu03,rugh13}.  Although several studies have modeled the \om-\oz\ relationship for varying \om\ abundances and different stellar hosts (e.g., \citealt{ratn72,levi79,kast80,kast85,segu03,greg21}), there has been no in-depth study evaluating the ability of \oz\ to predict \om\ as a biosignature.  In this series of papers, we will explore the \om-\oz\ relationship in depth for a variety of stellar hosts and atmospheric conditions.  For this first paper, we focus on the \om-\oz\ relationship for ``Earth-like'' planets for different stellar hosts.  Here we take Earth-like to mean a planet that has the same composition and size as Earth, receives the equivalent total incident flux from the Sun as modern Earth, and has a similar atmospheric composition.  This study currently contains the largest number of models run with a fully coupled climate and photochemistry code dedicated to understanding the \om-\oz\ relationship, along with exploring the widest range of stellar hosts as well as the largest number of different \om\ atmospheric abundances.

\begin{figure*}[h!]
\centering
\includegraphics[scale=0.55]{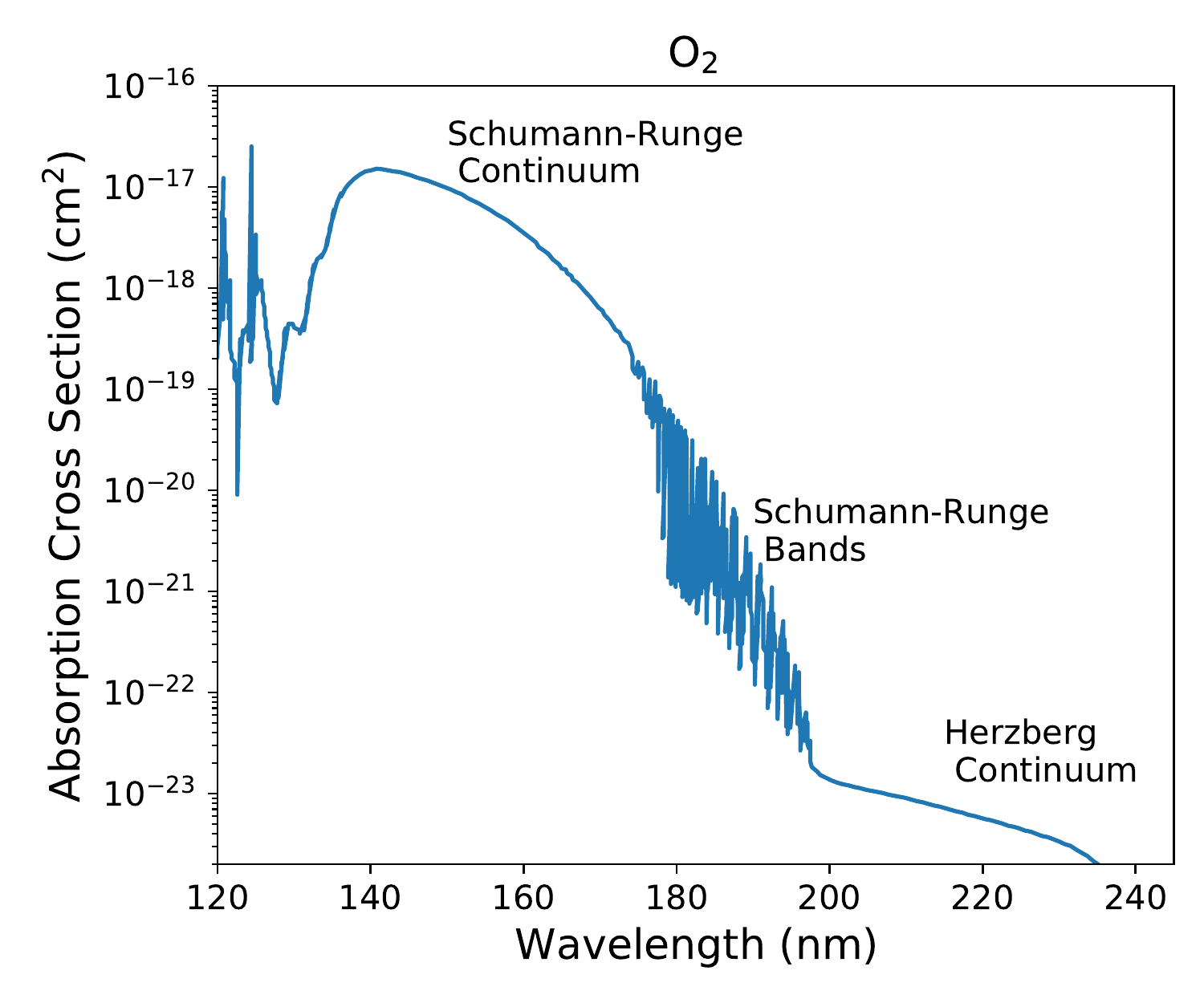}
\includegraphics[scale=0.55]{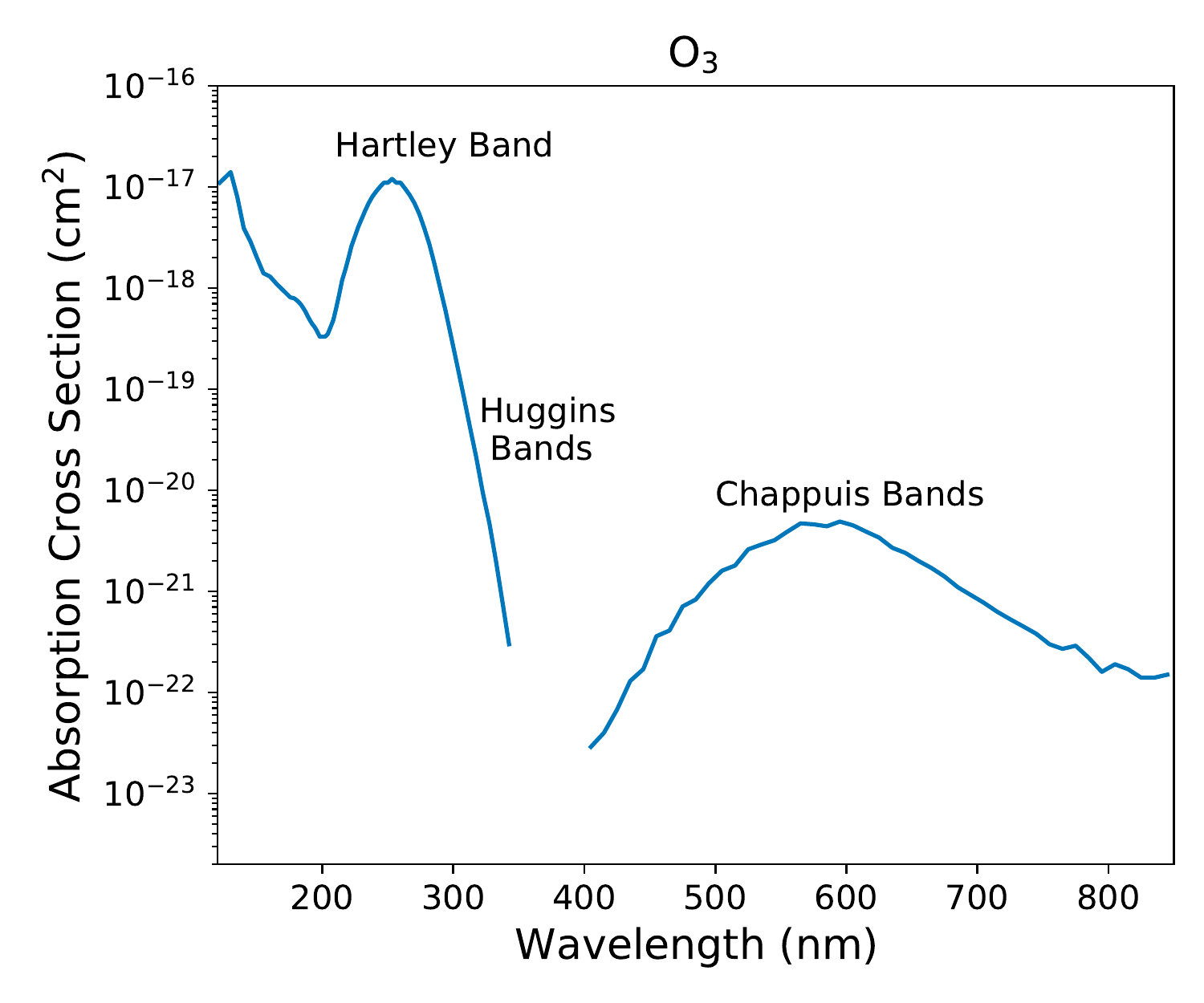}
\caption{Absorption cross sections of \om\ (left) and \oz\ (right), both plotted on the same y-axis scale to enable easier comparison.  The y-axis is the same for both to enable easier comparison. Relevant \om\ bands are the Schumann-Runge continuum (137-175 nm), the Schuman-Runge bands (175-200 nm), and the Herzberg continuum (195-242 nm). For less than 175 nm an excited O atom, the \od\ radical, is formed along with a ground state O atom during photolysis.  \oz\ bands are the Hartley bands (200-300 nm), the Huggins bands (310-350 nm), and the Chappuis bands (410-750 nm).  Photolysis within the Hartley bands will produce the \od\ radical while other \oz\ bands create a ground state O atom. Absorption cross section data is from \cite{brass05}. \label{fig:XS}}
\end{figure*}

\section{Chemistry of \oz\ production and destruction \label{sec:chem}}

In this section we  give a brief overview of the most important reactions for the production and destruction of \oz.  The wavelength-dependent absorption cross sections for \om\ and \oz\ are shown in Fig.~\ref{fig:XS} as a reference for the reader, as they determine photolysis rates in different wavelength regions. Incident stellar UV flux as well as the amount of nitrogen- and hydrogen-bearing species primarily control the concentration of \oz\ in the atmosphere.

\subsection{The Chapman mechanism}

Ozone is primarily created in the stratosphere by a set of reactions called the Chapman mechanism \citep{chap30}.   These reactions begin with the photolysis of \om,
\begin{equation}
\m{O}_2 + \m{h}\nu \rightarrow \m{O} + \m{O  (}175 < \lambda < 242\ \m{nm}),
\label{r:PO2_O}
\end{equation}
which creates ground state O atoms (also written as O($^3$P)), which are highly reactive due to two unpaired electrons.  These O atoms then combine with \om\ molecules to form \oz,
\begin{equation}
\m{O + O}_2 + M \rightarrow \m{O}_3 + M,
\label{r:O2M}
\end{equation}
where $M$ is a background molecule that carries away excess energy. Reaction~\ref{r:O2M} is a 3-body reaction, meaning it is more efficient at lower temperatures and higher atmospheric densities. It is faster in denser atmospheric regions with a larger availability of O atoms, causing bulk of \oz\ on Earth to exist in the stratosphere rather than at higher altitudes.

Photolysis of \om\ can also occur higher in the atmosphere with higher energy photons,
\begin{equation}
\m{O}_2 + \m{h}\nu \rightarrow \m{O }+ \m{O(}^1\m{D)  (}\lambda < 175\ \m{nm}),
\label{r:PO2_O1D}
\end{equation}
where the \od\ radical is created along with a ground state O atom.  Free radicals are by nature  extremely reactive as they have at least one unpaired valence electron.  Thus, they tend to have extremely brief lifetimes. The \od\ radical can return to the ground state by being ``quenched'' via a collision with a background molecule,
\begin{equation}
\m{O(}^1\m{D)} + M \rightarrow \m{O} + M ,
\label{r:quench}
\end{equation}
or it can react with another molecule.  Reactions with other molecules will be further explored in Sects.~\ref{sec:cc} and \ref{sec:chemresults}.

Although \om\ absorption cross sections are significantly larger at wavelengths that produce \od\ radicals ($<$ 175 nm; see Fig.~\ref{fig:XS}), these photons are absorbed high in the atmosphere and therefore do not contribute to the creation of stratospheric \oz\ on modern Earth.  Absorption caused by Lyman-$\alpha$ photons (121.6 nm) is generally absorbed in the mesosphere, and photons in the Schumann-Runge continuum (130-175 nm) are absorbed in the thermosphere.  Also note that some wavelengths shorter than Lyman-$\alpha$ can ionize \om, although that wavelength region is not included in our photochemistry model due to the low amount of photons in that region emitted by GKM stars (see Sect.~\ref{sec:atmos}).

Although \om\ photolysis from these higher energy photons ($<$ 175 nm) occurs above the stratosphere on modern Earth, planets with different amounts of atmospheric \om\ would experience absorption of these photons at varying atmospheric altitudes.  Less \om\ would allow high energy photons to travel deeper into the atmosphere before absorption via \om\ photolysis.  Although this will not cause the bulk of \oz\ formation, it will impact the upper atmospheric chemistry by creating more \od\ at lower altitudes.  The effects of this will discussed at length in Sect.~\ref{sec:chemresults}.

Once \oz\ is created by Reaction~\ref{r:O2M}, it is often quickly photolyzed.  \oz\ photolysis from a photon in the Hartley band (200-300 nm) will create an \od\ radical, while photons from the lower energy Huggins bands (310-350 nm), Chappuis bands (410-750 nm), and longer wavelengths will create a ground state O atom,
\begin{equation}
\m{O}_3 + \m{h}\nu \rightarrow \m{O}_2 + \m{O(}^1\m{D)}  (\lambda < \m{310\ nm}),
\label{r:PO3_O1D}
\end{equation}
\vspace{-0.7cm}
\begin{equation}
\m{O}_3 + \m{h}\nu \rightarrow \m{O}_2 + \m{O (310} < \lambda < \m{1140\ nm}).
\label{r:PO3_O}
\end{equation}
Photons with wavelengths shorter than 200 nm are often absorbed high in the atmosphere by \om\ and other molecules.  As with \om\ photolysis, \od\ radicals created by Reaction~\ref{r:PO3_O1D} will either be quenched by a background molecule and returned to the ground state (Reaction~\ref{r:quench}), or they will react with other molecules.

Photolysis of \oz\ is not seen as a loss of \oz, as the resulting O atom and \om\ molecule often quickly recombine into \oz\ via Reaction~\ref{r:O2M}.  Due to the rapid cycling between \oz\ and O, it is instead the conversion of \oz\ + O (called ``odd oxygen'') into \om\ that actually results in a loss of \oz, as occurs in the final reaction of the Chapman mechanism,
\begin{equation}
\m{O}_3 + \m{O} \rightarrow 2\m{O}_2.
\label{r:O3_O}
\end{equation}
Odd oxygen (\oz\ + O) being converted into \om\ is considered a loss of \oz\ because the photolysis of \om\ (Reaction~\ref{r:PO2_O},\ref{r:PO2_O1D}) is the slowest of the Chapman mechanism reactions, and the limiting factor in \oz\ production.  Therefore the loss of odd oxygen on long timescales causes a true decrease in \oz.

\subsection{Catalytic cycles of HO$_x$ and NO$_x$ \label{sec:cc}}

The Chapman mechanism on its own overestimates the amount of atmospheric \oz\ because it does not take into account catalytic cycles that destroy \oz.  These destruction cycles follow the format, 
\begin{equation*}
\begin{aligned}
\m{X + O}_3 \rightarrow \m{XO + O}_2 \\
\m{XO + O} \rightarrow \m{X + O}_2 \\
\hline
\m{Net:} \hspace{0.5cm}  \m{O}_3 + \m{O} \rightarrow 2\m{O}_2
\end{aligned}
\end{equation*}
where X is a free radical.  During this process  X and XO will cycle between each other while converting odd oxygen (\oz\ + O) into \om, similarly to the last step of the Chapman mechanism (Reaction~\ref{r:O3_O}).  As stated above, this results in the overall loss of \oz\ because \om\ photolysis is the limiting reaction of \oz\ formation.  X and XO can cycle between each other and continuously destroy \oz\ until reactions that convert either X or XO into non-reactive ``reservoir'' species occur. The primary catalytic cycles of \oz\ destruction in modern Earth's atmosphere are the HO$_x$ (hydrogen oxide) and NO$_x$ (nitrogen oxide) catalytic cycles.  We note that on modern Earth there are also \oz\ destroying catalytic cycles that are powered by molecular compounds primarily created anthropogenically (e.g.,\  chlorine and bromine cycles; \citealt{crut01}), but they will not be included in this study.

The HO$_x$ catalytic cycle is powered by the OH (hydroxyl) and HO$_2$ (hydroperoxyl) radicals.  When an \od\ radical is created either by photolysis of \om\ (Reaction~\ref{r:PO2_O1D}) or \oz\ (Reaction~\ref{r:PO3_O1D}) it can react with H$_2$O to form OH,
\begin{equation}
\m{H}_2\m{O} + \m{O(}^1\m{D)} \rightarrow \m{OH} + \m{OH}.
\label{r:H2O_OH}
\end{equation}
The OH radical is a major sink for multiple atmospheric gases (e.g., CH$_4$, CO) and is often called the `detergent of the atmosphere' for this reason.  It destroys \oz\ during the HO$_x$ catalytic cycle as follows,
\begin{equation}
\m{OH} + \m{O}_3 \rightarrow \m{HO}_2 + \m{O}_2,
\label{r:HOx_OH}
\end{equation}
\vspace{-0.6cm}
\begin{equation}
\m{HO}_2 + \m{O} \rightarrow \m{OH} + \m{O}_2.
\label{r:HOx_HO2}
\end{equation}
In addition to this primary destruction cycle, other HO$_x$ cycles can contribute significantly to \oz\ destruction via,
\begin{equation}
\m{OH} + \m{O}_3 \rightarrow \m{HO}_2 + \m{O}_2,
\tag{\ref{r:HOx_OH}}
\end{equation}
\vspace{-0.6cm}
\begin{equation}
\m{HO}_2 + \m{O}_3 \rightarrow \m{OH} + 2\m{O}_2,
\end{equation}
resulting in two \oz\ molecules converted to three \om\ molecules, or,
\begin{equation}
\m{OH} + \m{O} \rightarrow \m{H} + \m{O}_2,
\end{equation}
\vspace{-0.6cm}
\begin{equation}
\m{H} + \m{O}_2 + M \rightarrow \m{HO}_2 + M,
\label{r:HM}
\end{equation}
\vspace{-0.6cm}
\begin{equation}
\m{HO}_2 + \m{O} \rightarrow \m{OH} + \m{O}_2,
\tag{\ref{r:HOx_HO2}}
\end{equation}
with a net result of two O atoms converted into an \om\ molecule. Because OH production via \od\ is a byproduct of the Chapman mechanism, HO$_x$ catalytic cycle efficiency can be increased with higher rates of \oz\ formation.  This process can be slowed through reactions that convert OH/HO$_2$ into a reservoir species such as H$_2$O, HNO$_2$, or H$_2$O$_2$, which are significantly less reactive.

The NO$_x$ catalytic cycle destroys \oz\ with the NO (nitric oxide) and NO$_2$ (nitrogen dioxide) radicals.  The primary source of these radicals in the stratosphere is from N$_2$O (nitrous oxide) which is biologically produced by nitrification and denitrification processes within soil.  N$_2$O can additionally be produced anthropogenically, primarily through agriculture. It is converted into NO by interactions with the \od\ radical,
\begin{equation}
\m{N}_2\m{O} + \m{O(}^1\m{D)}    \rightarrow    \m{NO}    +  \m{NO}.
\label{r:N2O_NO} 
\end{equation}
A secondary source of NO is production via lightning in the upper troposphere, which can then be transported into the lower stratosphere.  The NO$_x$ catalytic cycle destroys O$_3$ as follows,
\begin{equation}
\m{NO} + \m{O}_3 \rightarrow \m{NO}_2 + \m{O}_2,
\label{r:NOx_NO}
\end{equation}
\vspace{-0.6cm}
\begin{equation}
\m{NO}_2 + \m{O} \rightarrow \m{NO} + \m{O}_2.
\end{equation}
NO$_x$  can destroy \oz\ with the following cycle as well,
\begin{equation}
\m{NO} + \m{O}_3 \rightarrow \m{NO}_2 + \m{O}_2,
\tag{\ref{r:NOx_NO}}
\end{equation}
\vspace{-0.6cm}
\begin{equation}
\m{NO}_2 + \m{O}_3 \rightarrow \m{NO}_3 + \m{O}_2.
\end{equation}
\vspace{-0.6cm}
\begin{equation}
\m{NO}_3 + \m{h}\nu \rightarrow \m{NO} + \m{O}_2.
\end{equation}
with a net conversion of two \oz\ molecules into three \om\ molecules.  NO$_x$ reactions are highly temperature dependent and are faster at hotter temperatures.  The main reservoir species associated with NO$_x$ are HNO$_3$ and N$_2$O$_5$, which have slow photolysis rates.

We note that although in the stratosphere NO$_x$ destroys \oz\ through this catalytic cycle, that lower in the atmosphere it can help create \oz\ through the ``smog mechanism'' (see Sect.~\ref{sec:o2o3relationship}).  This low altitude \oz\ is a pollutant that can cause biological damage.  In this study we will focus on the majority of \oz\ in the stratosphere created by the Chapman mechanism.  In this study we will focus on the efficiency of the Chapman mechanism, along with the ability of the HO$_x$ and NO$_x$ catalytic cycles to destroy \oz\ for varying \om\ levels around different host stars.

\section{Methods \label{sec:methods}}

\subsection{Atmospheric models \label{sec:atmos}}

We modeled planetary atmospheres with \texttt{Atmos}\footnote{https://github.com/VirtualPlanetaryLaboratory/atmos}, a 1D coupled climate and photochemistry code to explore \oz\ formation for varying levels of \om\ on Earth-like planets around a variety of host stars.  Numerous studies have used either of these climate or photochemistry modules, as well as both coupled (e.g., \citealt{arne17,mead18a,linc18L,madd20,greg21,teal22}).  We give a brief overview of \texttt{Atmos} and refer readers to \cite{arne16} and \cite{mead18a} for extensive details.

\begin{figure*}[h!]
\centering
\includegraphics[scale=0.55]{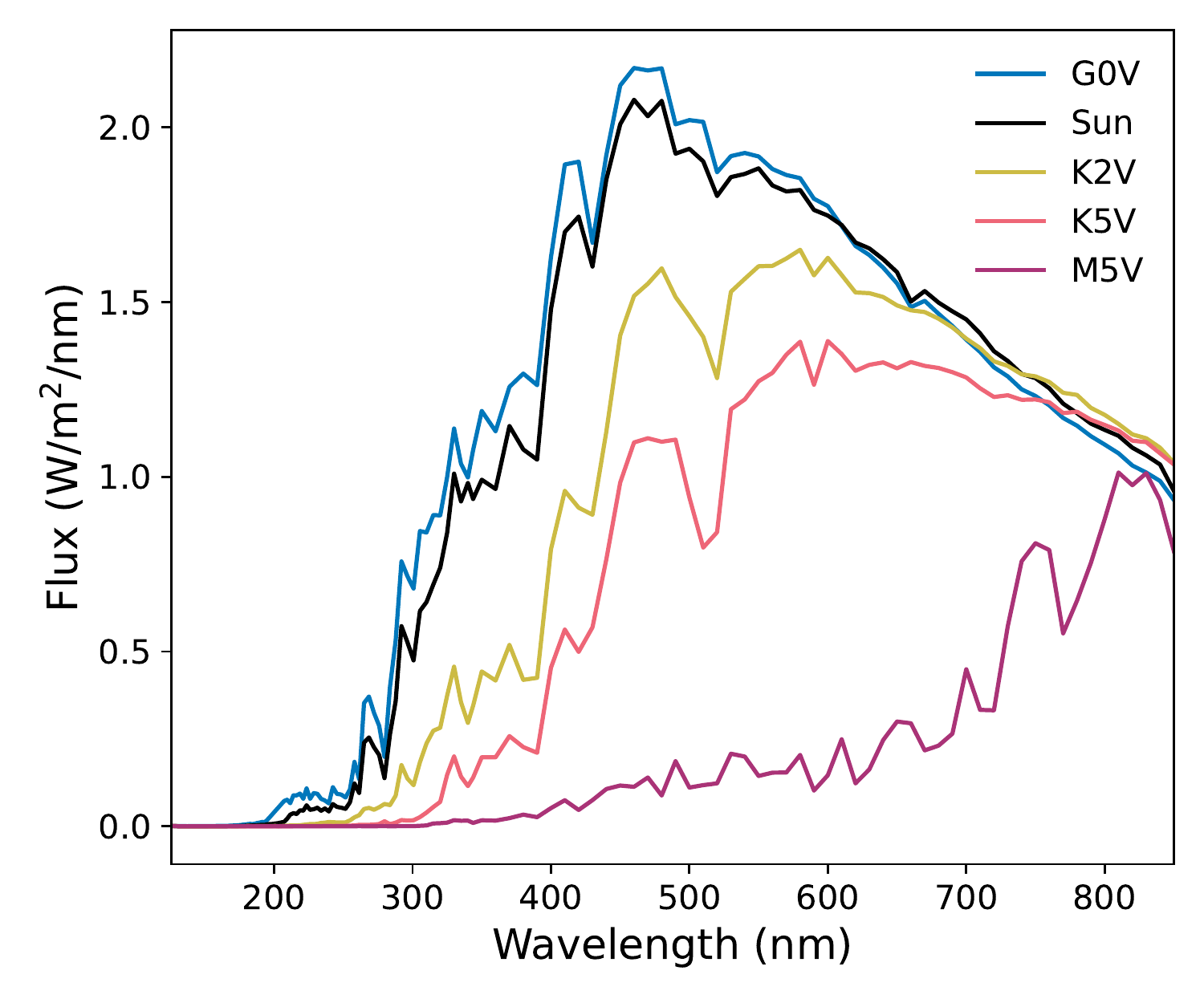}
\includegraphics[scale=0.55]{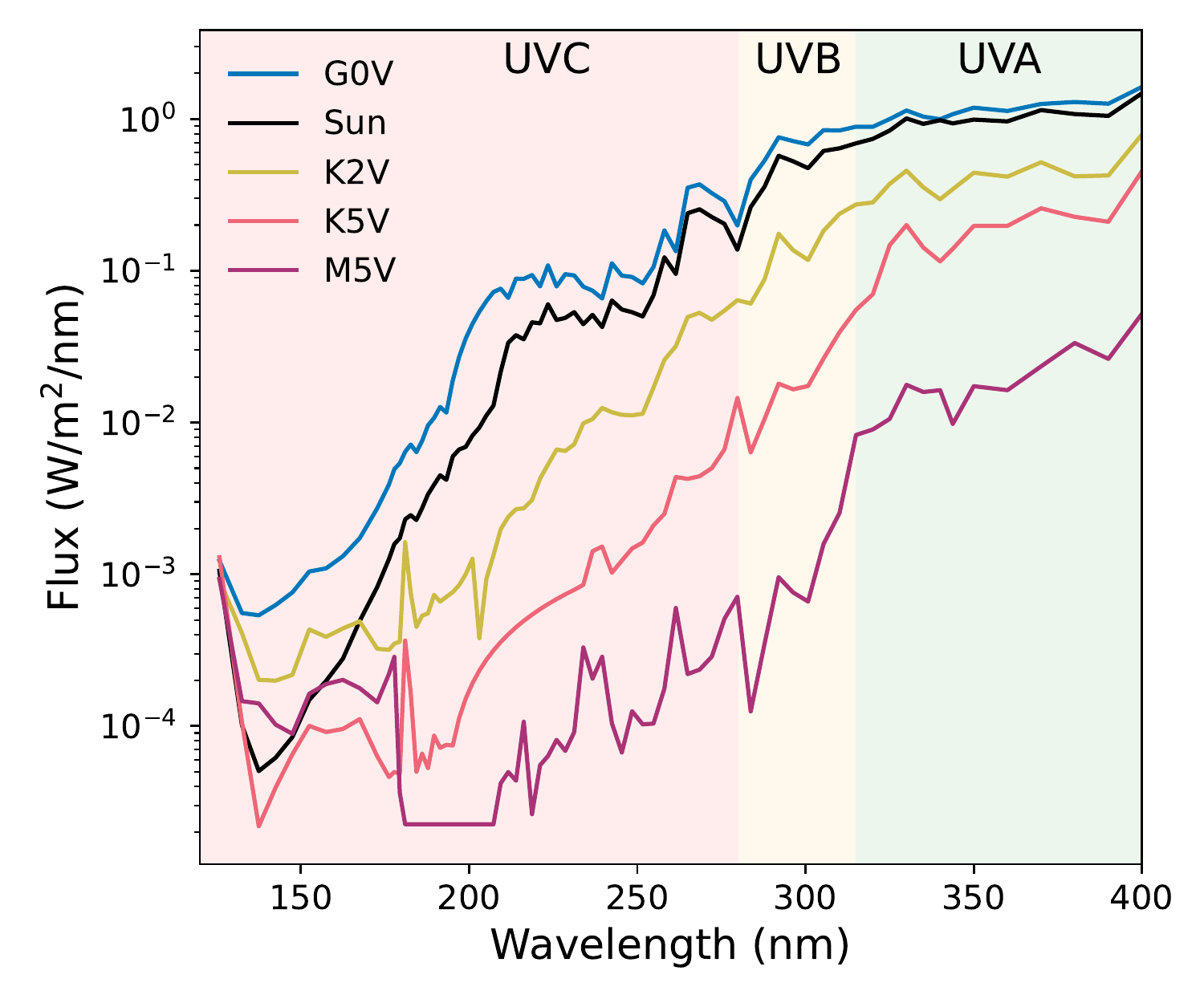}
\caption{Stellar spectra of the planet host stars.  G0V-K5V hosts are comprised of IUE UV data from \cite{rugh13}, and the M5V host comes from UV observations of GJ 876 from the MUSCLES survey \citep{fran16}.  Visible and IR wavelengths use ATLAS models of the same stellar temperature \citep{kuru79}.  The right-hand figure zooms in on the UV region relevant for photolysis with important biological UV regimes (see Sect.~\ref{sec:UV}).
\label{fig:stellarspectra}}
\end{figure*}

The photochemistry model  originates from \cite{kast79} and was expanded upon and updated by \cite{zahn06}.  It has been used extensively by many studies (e.g.,\ \citealt{kast80,segu03,segu05,segu10,doma14,greg21}).  The atmosphere is broken up in 200 plane parallel layers from 0 to 100 km.  The abundance of each gaseous species is calculated simultaneously with the flux and continuity equations using a reverse-Euler method for individual atmospheric layers.  Vertical transport between different layers include molecular and eddy diffusion.   Radiative transfer is computed with a $\delta$-2-stream method as described in \cite{toon89}.  For modern Earth \texttt{Atmos} uses 50 gaseous species, with nine of them being short lived and thus not included in transport calculations.  The photochemistry model is considered converged when its adaptive time step length reaches 10$^{17}$ seconds within 100 time steps.

The climate model was originally developed by \cite{kast86}, but has been significantly updated as described in \cite{kopp13} and \cite{arne16}.  Multiple studies have used this code to calculate habitable zones around a variety of stellar hosts used to study habitable zones and atmospheres of Earth-like planets around different stars (e.g.,\  \citealt {kopp13,segu03,segu05,segu10}).  The atmosphere is broken up into 100 plane parallel layers from the surface to an atmospheric pressure of 1 mbar.  A correlated-$k$ method computes the absorption of \oz, H$_2$O, CH$_4$, CO$_2$, and C$_2$H$_6$ throughout the atmosphere.  Total absorption of incident stellar flux is calculated for each atmospheric layer with a $\delta$-2-stream scattering algorithm \citep{toon89}, and outgoing IR radiation is calculated with correlated-$k$ coefficients for each species individually.  Updated H$_2$O cross sections from \cite{ranj20} have been incorporated into the code.  Convergence criteria are reached when both changes in temperature and the flux out of the top of the atmosphere are sufficiently small ($<$10$^{-5}$).

We run the climate and photochemistry models coupled with  inputs including host stellar spectrum (121.6 - 45450 nm), initial mixing ratios of atmospheric species, upper and lower boundary conditions for individual species, and initial temperature/pressure profiles.  Using initial conditions the photochemistry code  runs first and then transfer computed H$_2$O, CH$_4$, CO$_2$, and C$_2$H$_6$ mixing ratio profiles to the climate code.  The climate code  then updates the temperature and H$_2$O vapor profiles to feed back into the photochemistry.  These processes iterate with profiles from the photochemistry allowing for more accurate climate code calculations and vice-versa, until a converged solution is reached.

The climate code has not been successfully run to convergence for the same atmospheric height as the photochemistry code \citep{arne16}, so temperature and H$_2$O profiles of the upper, thin part of the atmosphere (typically $<$ 60-70 km) are held constant at the highest computed value from the climate code.  Sensitivity tests from \cite{arne16} suggest that the impact on the radiative transfer and climate of these models is not significant.

This study also implements the ``short-stepping'' method of convergence, as described in \cite{teal22}.  When iterating back and forth between the photochemistry and climate code, occasionally the code will oscillate between two different solutions.  For example, if the photochemistry code computes a large quantity of \oz, the climate code will respond with a large amount of atmospheric heating.  However, due to the temperature sensitivity of \oz\ production (Reaction~\ref{r:O2M}), this hotter atmosphere will cause lower amounts of \oz\ on the subsequent photochemistry iteration.  Using the ``short-stepping'' method we do not allow the climate code to fully adjust to the updated atmospheric profiles from  the photochemistry on a single iteration, and instead reach convergence slowly by iterating back and forth between the climate and photochemistry codes until a stable solution is reached.

\begin{table}[t!]
\begin{center}
\label{tab:stars}
\caption{Stellar hosts}
\begin{tabular}{crrr}
\hline 
\hline
Host Star &  Model &  FUV/NUV & UV data \\
 & T$_{\m{\scriptsize eff}}$ (K) &  &  source \\
\hline
G0V 	 & 6000 	 & 0.0028	 & a \\
Sun 	 & 5800 	 & 0.0010	 & a \\
K2V 	 & 5000 	 & 0.0010	 & a \\
K5V 	 & 4500 	 & 0.0012	 & a \\
M5V 	 & 3000 	 & 0.0084	 & b \\
\hline
\hline
\end{tabular}
\end{center}
\small{
a - \cite{rugh13}\\
b - \cite{fran16}
}
\end{table}

We modeled planetary atmospheres orbiting a variety of stellar hosts (see Sect.~\ref{sec:stellarspectra}) at the Earth-equivalent distance with varying levels of \om.  Here we take Earth-equivalent distance to mean that the planet receives the same total amount of incident flux from their parent stars as modern Earth receives from the Sun.  We set \om\ as a constant mixing ratio for all cases, with values varying from 0.01-150\% PAL \om\ (mixing ratios of 2.1$\times10^{-3}$ - 0.315).  Higher \om\ levels are not explored because large \om\ levels would unstable with biological compounds \citep{kump08}. Lower \om\ are not modeled because it is thought that \om\ abundances from $\sim10^{-3}$\% to $\sim$1\% PAL are not expected to be stable in an Earth-like atmosphere as calculated by \cite{greg21} (details on these limits in Sect.~\ref{sec:O3biosignature}).

Other initial conditions for the models were chosen to resemble modern Earth including atmospheric mixing ratios, planetary composition, and size.  \texttt{Atmos} haze production was not used.  All models were run at a zenith of 60$^\circ$ degrees (Lambertian average) and with cloudless skies.  Fixed mixing ratios were used for CH$_4$ (1.8$\times10^{-6}$), N$_2$O (3.0$\times10^{-7}$), and CO$_2$ (3.6$\times10^{-4}$).  All other species used initial atmospheric profiles and boundary conditions as defined in \texttt{Atmos}'s modern Earth template, and surface pressure remained constant at 1~bar.  We note  that defining CH$_4$ at a constant mixing ratio resembling modern Earth differs from several studies modeling ``Earth-like'' planets, which have adjusted CH$_4$ mixing ratios to reflect the CH$_4$ ground flux of modern Earth, resulting in much higher atmospheric CH$_4$ mixing ratios (e.g., \citealt{rugh15,wund19,teal22}).  We chose to maintain  CH$_4$ mixing ratio of modern Earth to better isolate the effects of different stellar hosts on the \om-\oz\ relationship.  The impact of changing CH$_4$ levels on \oz\ abundance is discussed further in Sect.~\ref{sec:o2o3relationship}.

\subsection{Input stellar spectra \label{sec:stellarspectra}}

All host star spectra inputted into \texttt{Atmos} comprise of actual UV observations supplemented with synthetic ATLAS model spectra \citep{kuru79} for the visible and IR. Table~\ref{tab:stars} contains information about the host stars and their spectra are shown in Fig.~\ref{fig:stellarspectra}. The G0V-K5V stellar spectra were created in \cite{rugh13} and are a combination of UV data from the \emph{International Ultraviolet Explorer} (IUE) data archives\footnote{http://archive.stsci.edu/iue} and model ATLAS spectra for the same stellar temperature \citep{kuru79}.  UV data for the M5V host is from GJ 876 observations obtained by the \emph{Measurements of the Ultraviolet Spectral Characteristics of Low-mass Exoplanetary Systems} (MUSCLES) survey \citep{fran16}.

The UV spectrum of a planet's host star is extremely important in the photochemical modeling of \oz\ production.  Not only does the total amount of UV dictate photolysis rates, but the UV spectral slope determines the creation and destruction rates of \oz.  The far-UV (FUV; $\lambda <$ 200 nm) is primarily responsible for photolysis of \om\ (and the creation of \oz), while the mid- and near-UV (abbreviated NUV, for brevity; 200 nm $< \lambda <$ 400 nm) is responsible for the photolysis of \oz.  The NUV additionally can photolyze H$_2$O, which creates the HO$_x$ species responsible for destroying \oz, causing NUV flux to destroy \oz\ both directly and indirectly.  Hence, a higher FUV/NUV flux ratio will create \oz\ more efficiently.  Low-mass, active stars tend to have lower FUV/NUV flux ratios as activity will cause excess FUV chromospheric radiation, while NUV wavelengths are often absorbed for cool stars by TiO \citep{harm15}.

\subsection{Radiative transfer model}

After \texttt{Atmos} computes the compositions of our model atmospheres, the \emph{Planetary Intensity Code for Atmospheric Scattering Observations} (\texttt{PICASO}) computes planetary emission spectra \citep{bata19,bata21}.  \texttt{PICASO} is a publicly available\footnote{https://natashabatalha.github.io/picaso/index.html} radiative transfer code capable of producing transmission, reflected light, and emission spectra for a diverse range of planets.  Our emission spectra were calculated at a phase angle of 0$^\circ$ (full phase) with altitude dependent pressure, temperature and mixing ratio profiles computed by \texttt{Atmos}.  Output spectra cover a wavelength range of 0.3 - 14 $\mu$m, although particular focus is put on the \oz\ 9.7 $\mu$m feature in this study.

\begin{figure*}[h!]
\centering
\includegraphics[scale=0.55]{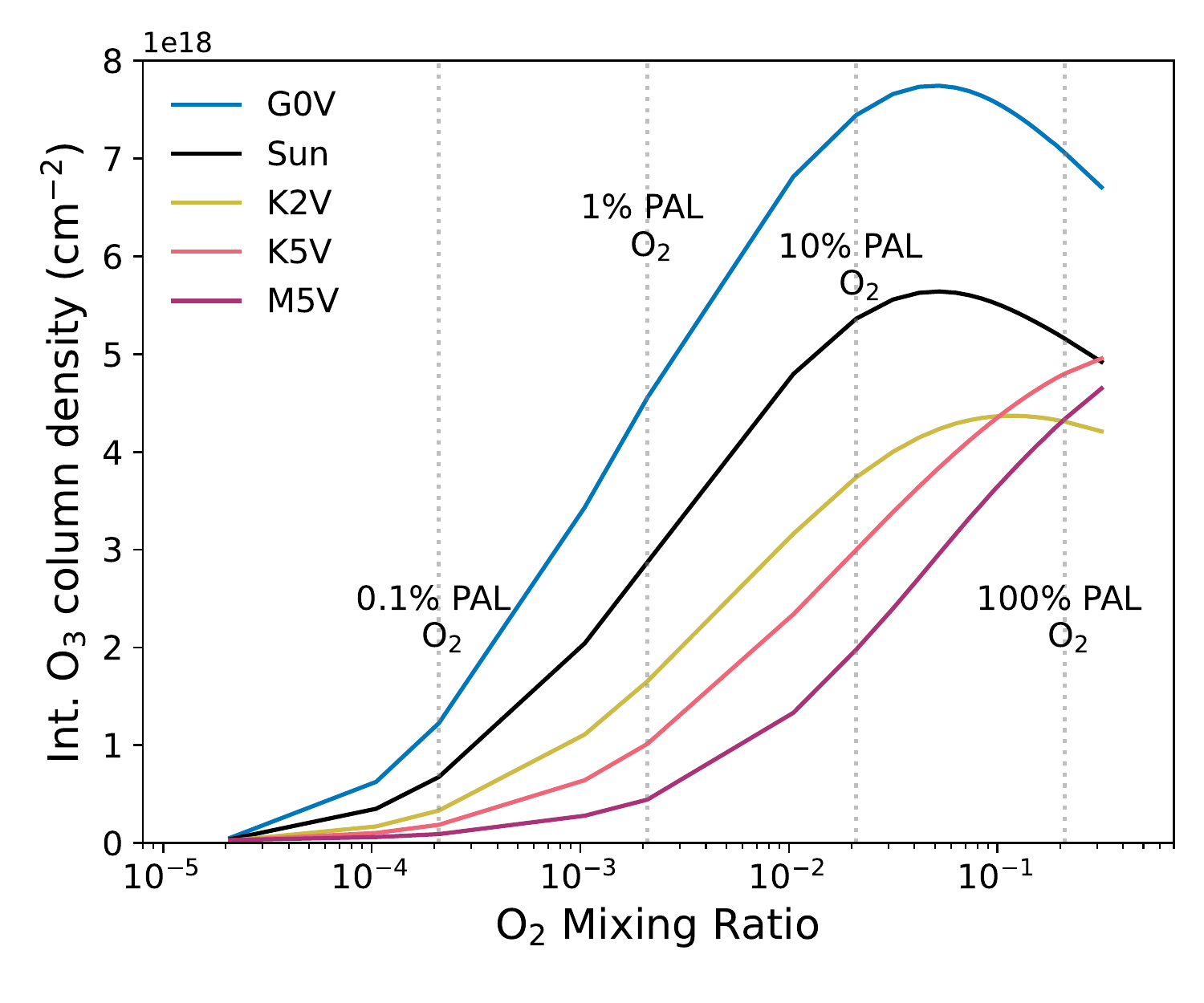} 
\includegraphics[scale=0.55]{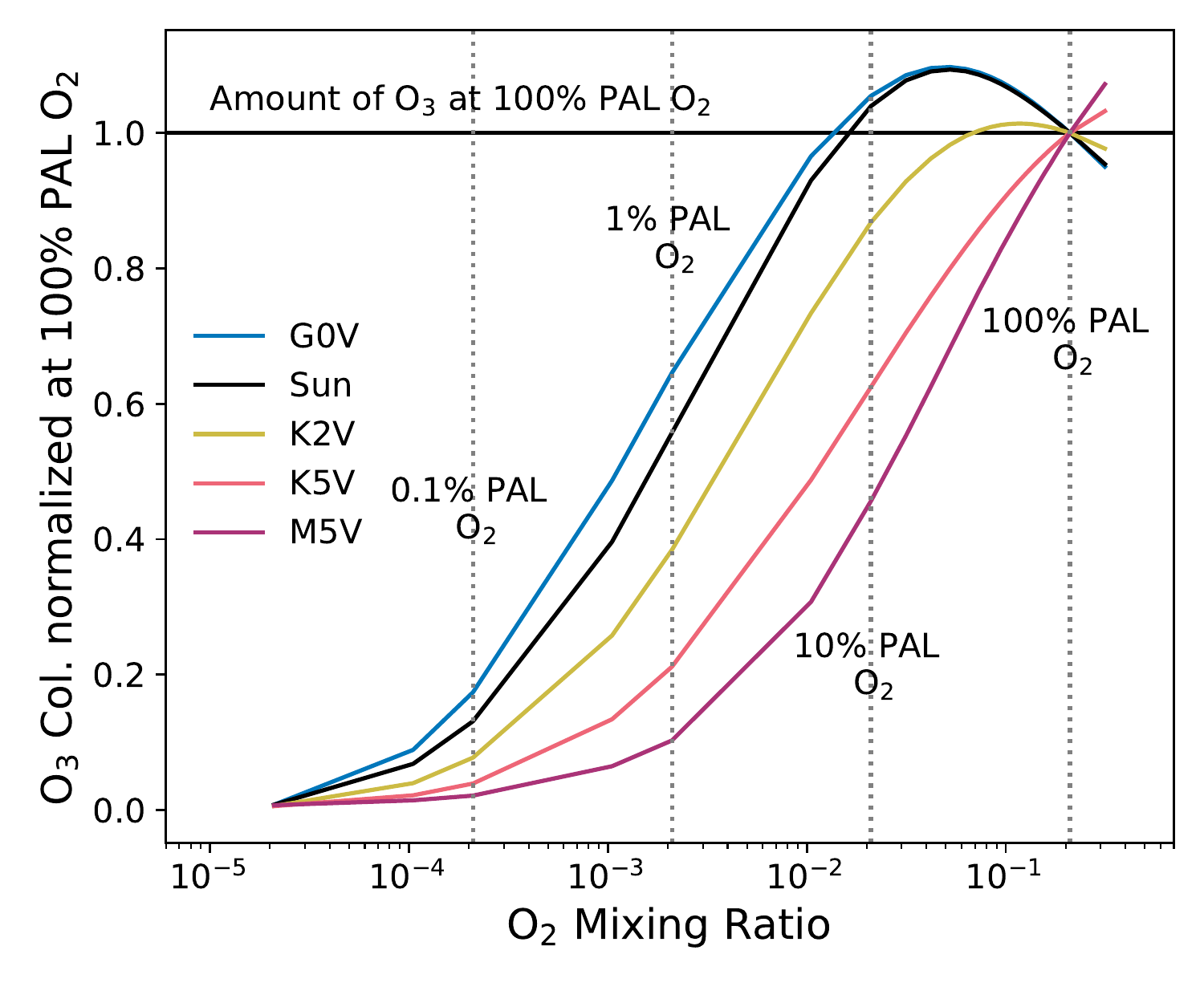}
\caption{O$_2$-O$_3$ relationship for Earth-like planets orbiting different host stars (indicated in legend) with varying amounts of O$_2$ mixing ratios. Molecular oxygen levels compared to Earth's present atmospheric level (PAL) are shown as dashed vertical lines for reference.  
The left-hand figure shows the relationship in terms of total integrated \oz\ column density, and the right-hand figure in terms of total \oz\ normalized at the amount produced in the 100\% PAL \om\ case for each stellar host. The nonlinearity in these relationships is primarily due to the pressure dependence of  Reaction~\ref{r:O2M} which forms O$_3$.  The main takeaway is that the \om-\oz\ relationship is significantly different for hotter host stars (G0V, Sun, K2V) versus cooler host stars (K5V, M5V), with hotter hosts experiencing peak \oz\ formation occurring at \om\ levels under 100\% PAL. See Sect.~\ref{sec:chemresults} for a detailed explanation. \label{fig:O3-O2_MR_all}}
\end{figure*}

\section{Results}

\subsection{\om-\oz\ relationship \label{sec:chemresults}}

Figure~\ref{fig:O3-O2_MR_all} shows the \om-\oz\ relationship for all of our model planetary atmospheres.  The \om-\oz\ relationship is highly dependent on the stellar host, with different trends for model atmospheres having  hotter host stars (G0V, Sun, K2V) versus cooler host stars (K5V, M5V). Since \oz\ is produced via the Chapman mechanism by converting \om\ into \oz, one would naively expect the \oz\ concentration to increase as the \om\ mixing ratio increases, which  the case for the cooler host stars. However, the \om-\oz\ relationship for hotter stellar hosts behaves unexpectedly such that \oz\ abundance peaks and then decreases as the abundance of \om\ decreases from modern Earth levels.  Maximum \oz\ abundance occurs in the 25\% PAL \om\ models for  the G0V and Sun hosts, and the 55\% PAL \om\ model for the K2V host.  This effect does not occur for cooler host stars, with \oz\ abundance dropping consistently for models with less \om, though not in a linear fashion.  As a result, the K5V and M5V models with maximum \om\  considered (150\% PAL) created the maximum amount of \oz.  These results are summarized in Table~\ref{tab:maxO3}.  To allow for a simple parameterization of the \om-\oz\ relationships shown in Fig.~\ref{fig:O3-O2_MR_all} that can be used as an approximation in, for instance, GCM and retrieval modeling, we fit a fifth degree polynomial of the form, 
\begin{equation}
    y = ax^5 + bx^4 + cx^3 + dx^2 + ex + f,
\end{equation}
where $y$ is the integrated \oz\ column density (cm$^{-2}$), $x$ is the base 10 logarithm of the O$_2$ mixing ratio, and $a$, $b$, $c$, $d$, $e$, and $f$ are the best fit polynomial coefficients listed in Table~\ref{tab:coefs}.  This fit is valid over the range of \om\ abundances modeled in this study (0.01\%-150\% PAL).

\begin{table}[t!]
\begin{center}
\caption{Maximum  Integrated O$_3$ Column Density
\label{tab:maxO3}}
\begin{tabular}{c|r|r}
\hline 
\hline
Host Star &  Max Int. \oz\ Col. & Max \oz\ Model  \\
 & Density (10$^{18}$ cm$^{-2}$) & (\% PAL \om) \\
\hline
G0V	& 7.74	 & 25 \\
Sun	& 5.64	 & 25 \\
K2V	& 4.37	 & 55 \\
K5V	& 4.96	 & 150 \\
M5V	& 4.65	 & 150 \\
\hline
\hline
\end{tabular}
\end{center}
\end{table}

\begin{table*}[h!]
\begin{center}
\caption{Coefficients of polynomial fit of the \om-\oz\ relationship
\label{tab:coefs}}
\begin{tabular}{lcccccc}
\hline 
\hline
Host Star & a & b & c & d & e & f\\
\hline
G0V 	 & 4.582e+16	 & 6.021e+17	 & 2.481e+18	 & 2.618e+18	 & -1.126e+18	 & 5.742e+18 \\
Sun 	 & 6.203e+16	 & 7.716e+17	 & 3.147e+18	 & 4.157e+18	 & 8.412e+17	 & 4.640e+18 \\
K2V 	 & 3.417e+16	 & 3.846e+17	 & 1.283e+18	 & 7.391e+17	 & -8.266e+17	 & 3.734e+18 \\
K5V 	 & 6.826e+15	 & 2.379e+16	 & -3.424e+17	 & -1.848e+18	 & -9.440e+17	 & 4.898e+18 \\
M5V 	 & -2.144e+16	 & -3.381e+17	 & -1.933e+18	 & -4.458e+18	 & -1.941e+18	 & 4.542e+18 \\
\hline
\hline
\end{tabular}
\end{center}
\end{table*}

\begin{figure*}[h!]
\centering
\includegraphics[scale=0.35]{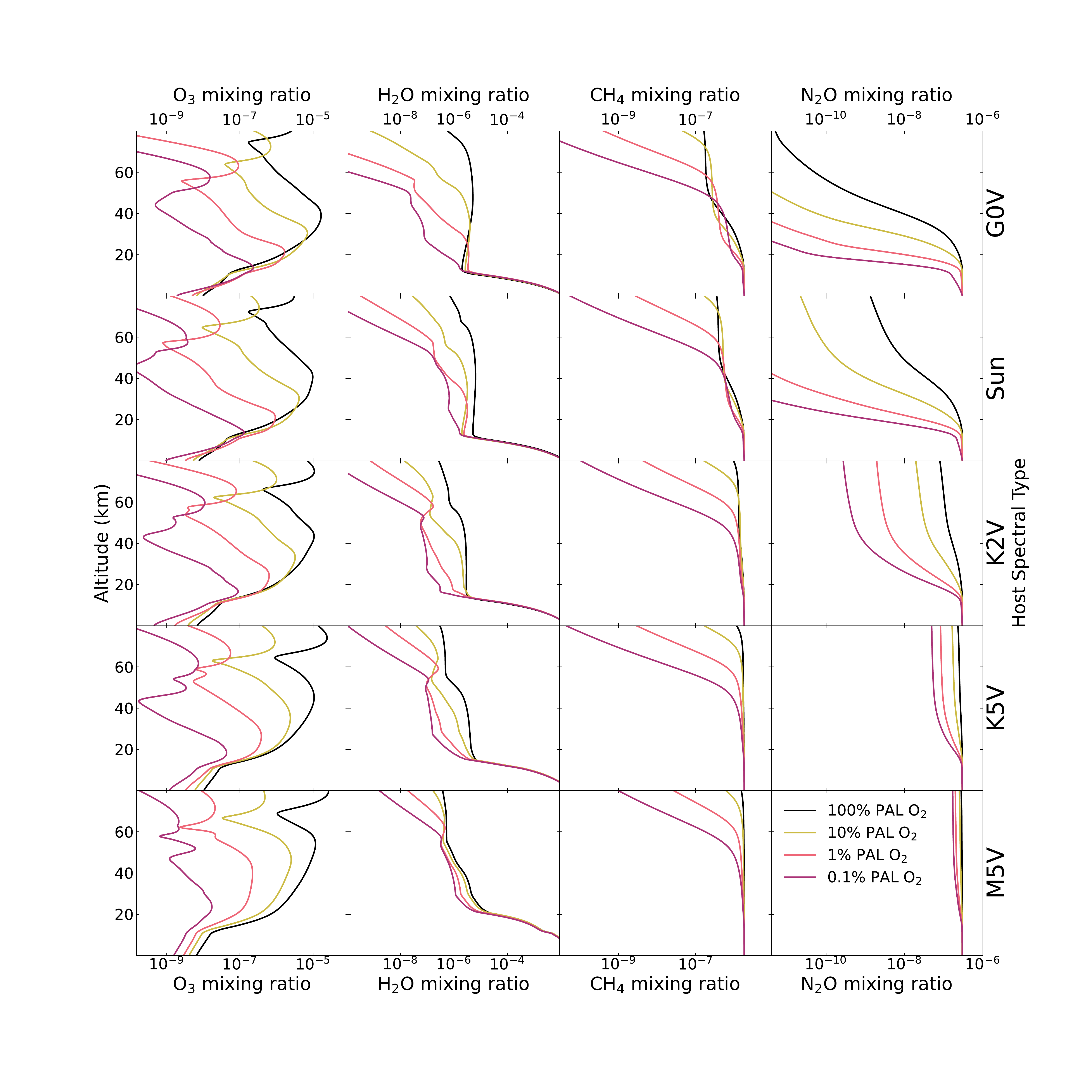}
\vspace{-1cm}
\caption{Mixing ratio profiles of O$_3$, H$_2$O, CH$_4$, and N$_2$O; all potential biosignature species.  Each row represents results for model atmospheres orbiting different stellar hosts as indicated on the right-hand vertical axis.  Each plot shows the mixing ratio profiles for \om\ abundances of 100\%, 10\%, 1\% and 0.1\% PAL.  \om\ on modern Earth is a significant UV shield, allowing model atmospheres with decreasing \om\ values to have photolysis occur at consistently lower altitudes.  This is demonstrated well with the bulk of \oz\ forming at lower altitudes with less \om, as well as increased upper atmospheric depletion for H$_2$O, CH$_4$, and N$_2$O via  photolysis. The downward shift of \oz\ and upper atmospheric depletion of other species is shown to decrease for model atmospheres around cooler stars with lower incident UV flux, and hence lower photolysis rates.  Full details of this atmospheric chemistry is shown in Sects.~\ref{sec:chemresults} and \ref{sec:catalyticresults}.  \label{fig:genchem}}
\end{figure*}

The seemingly counterintuitive phenomenon of hotter hosts having \oz\ levels \textit{increase} as \om\ levels \textit{decrease} can be explained by two factors: UV shielding abilities of \om, and the pressure dependency of \oz\ formation.  First, we will address the UV shielding ability of \om.  Despite the fact that \om\ UV absorption cross sections are either significantly smaller than those of \oz\ or require far higher energy photons (see Fig.~\ref{fig:XS}), \om\ remains an important UV shield on modern Earth, primarily due to its large abundance.  Although \om\ is less efficient at absorbing UV photons than \oz, \om\ makes up $\sim$21\% of the atmosphere, whereas \oz\ is a trace gas with a maximum value of $\sim$10 ppm on modern Earth. This allows the far larger number of \om\ molecules to compensate for its smaller absorption cross-sections and absorb many photons with wavelengths shorter than 240~nm (the required wavelength for \om\ photolysis, see Reaction~\ref{r:PO2_O}).  As a result, as \om\ decreases, UV shielding in that wavelength range decreases, allowing photolysis to occur deeper in the atmosphere.  This is illustrated in Fig.~\ref{fig:genchem}, where mixing ratio profiles of \oz, H$_2$O, CH$_4$, and N$_2$O are shown for all host stars at \om\ abundances of 100\%, 10\%, 1\%, and 0.1\% PAL \om.  Photolysis occurs at lower atmospheric altitudes  as \om\ decreases, leading the \oz\ layer to shift downward in the atmosphere.  This effect is more pronounced for hotter host stars with high UV fluxes (particularly high FUV fluxes capable of \om\ photolysis), and therefore higher photolysis rates. 

Secondly, the depth in which \om\ photolysis occurs is of particular importance to the altitudes at which the \oz\ forming Chapman mechanism takes place, because as \om\ decreases, photolysis reaches not only deeper but also denser regions of the atmosphere.  This is of significant relevance to \oz\ formation due to the pressure dependency of the Chapman mechanism: Reaction~\ref{r:O2M}, in which an O atom and \om\ molecule combine (with the help of a background molecule) to form \oz, is a 3-body reaction, and therefore is faster at higher atmospheric densities.  Denser regions allow O, \om, and background molecules to come together and react more rapidly than in a thinner region of the atmosphere. For hotter host stars in our sample (G0V, Sun, K2V), the UV fluxes are strong enough to allow \om\ photolysis to reach much denser atmospheric layers as \om\ decreases, allowing the benefit of faster \oz\ production via Reaction~\ref{r:O2M} to outweigh the smaller source of \om, resulting in peak \oz\ abundance at lower \om\ levels.  Our cooler host stars (K5V, M5V), however, have weaker UV fluxes, meaning photons capable of \om\ photolysis cannot travel as deep in the atmosphere as for hotter hosts when \om\ decreases.  The additional speed of the Chapman mechanism for lower \om\ does not make up for the smaller amount of \om, causing \oz\ abundance to decrease for decreasing \om\ abundance with these cooler host stars.

This result of an increase in \oz\ production as \om\ levels decrease has been noted for the Earth-Sun system by several studies (e.g., \citealt{ratn72,levi79,kast80,kast85,lege93}), although this is the first time it has been explored for Earth-like planets around different stellar hosts.  Whether or not this would occur for an Earth-like planet will depend on if the UV flux (particularly the FUV flux) from its host star will be strong enough to incite \om\ photolysis at dense enough atmospheric levels that the increased rate of \oz\ production will be enough to counter the decreased amounts of \om\ from which \oz\ can form.  This effect contributes to the strong dependency of the \om-\oz\ relationship on the spectral type of the host star.  For example the G0V host star models have more \oz\ at 10\% PAL \om\ than at 100\% PAL \om, whereas for the M5V host star \oz\ abundance in the 10\% PAL \om\ model is nearly 60\% percent less than it is for the 100\% PAL \om\ model.

When looking at specific \oz\ mixing ratios, Fig.~\ref{fig:O3-O2_MR_all} also indicates an increase in \oz\ above the stratosphere for all models.  This upper atmosphere \oz\ (the ``secondary \oz\ layer'') is produced primarily by \om\ photolysis from higher energy photons ($>$ 175 nm; Reaction~\ref{r:PO2_O1D}) which produces the radical \od.  Photons of these wavelengths are absorbed high in the atmosphere but do not contribute significantly to stratospheric \oz, even when a decrease in \om\ allows photolysis to reach deeper layers of the atmosphere.  Instead, these photons create \oz\ above the primary \oz\ layer, generally in the mesosphere and thermosphere (see Sect.~\ref{sec:chem} for more details). Although \oz\ mixing ratios are high at these altitudes, due to the thin atmosphere, \oz\ creation at these elevations does not add considerably to the total amount of \oz.

Also note that although the K2V host star produces enough UV photons capable of \om\ photolysis to have peak \oz\ production in its 55\% PAL \om\ model, both the K5V and M5V host models show a larger amount of \oz\ than the K2V host for \om\ levels near 100\% PAL (Fig.~\ref{fig:O3-O2_MR_all}).  This is because although the K2V host has more FUV than the K5V and M5V hosts (Fig.~\ref{fig:stellarspectra}), the cooler hosts have higher FUV/NUV ratios (Table~\ref{tab:stars}), allowing more efficient \oz\ production without as much NUV \oz\ destruction.

In summary, the \om-\oz\ relationship is highly dependent on the UV flux of the host star, with different trends for hotter and cooler host stars.  Hotter host stars with high FUV fluxes experience peak \oz\ abundance at lower \om\ levels due to \oz\ formation occurring in deeper, denser parts of the atmosphere where the Chapman mechanism is more efficient.  Cooler host stars do not emit enough FUV flux for this effect to occur, and experience consistently decreasing \oz\ as \om\ decreases.

\subsection{Impact of varying \om\ on H$_2$O, CH$_4$, \& N$_2$O}

Figure~\ref{fig:genchem} shows the impact of varying \om\ levels on the biologically relevant atmospheres species H$_2$O, CH$_4$, and N$_2$O.  As \om\ decreases, photons usually absorbed by \om\ ($\lambda <$ 240 nm) travel deeper into the atmosphere and drive the majority of atmospheric changes.  This allows photolysis in general to reach lower altitudes, as well as photolysis caused by high energy photons that create the \od\ radical, which reacts quickly with many species.  \od\ is produced via photolysis of \om, \oz, N$_2$O, and CO$_2$ as follows,
\begin{equation}
\m{O}_2 + \m{h}\nu \rightarrow \m{O }+ \m{O(}^1\m{D)  (}\lambda < 175\ \m{nm}),
\tag{\ref{r:PO2_O1D}}
\end{equation}
\vspace{-0.7cm}
\begin{equation}
\m{O}_3 + \m{h}\nu \rightarrow \m{O}_2 + \m{O(}^1\m{D)  (}\lambda < 310\ \m{nm}),
\tag{\ref{r:PO3_O1D}}
\end{equation}
\vspace{-0.7cm}
\be
\m{N}_2\m{O} + \m{h}\nu \rightarrow \m{N}_2 + \m{O(}^1\m{D) (}\lambda < 200 \m{ nm)},
\label{r:PN2O_O1D}
\ee
\vspace{-0.7cm}
\be
\m{CO}_2 + \m{h}\nu \rightarrow \m{CO} + \m{O(}^1\m{D) (}\lambda < 167 \m{ nm)}.
\label{r:PCO2_O1D}
\ee
As \om\ decreases, \od\ creation moves deeper into the atmosphere for all these species.  For our models, \oz\ photolysis consistently creates the most \od, particularly at lower atmospheric heights.  \om\ photolysis is also a significant producer of \od, although it is limited to the stratosphere and above, even for the lowest \om\ levels modeled in this study.  CO$_2$ and N$_2$O photolysis contribute to \od\ production as well, although CO$_2$ photolysis is constrained to the upper atmosphere similarly to \om\ photolysis, while N$_2$O photolysis can occur much closer to the planetary surface for low \om\ levels.  Increased rates of photolysis as \om\ shielding decreases as well as increased \od\ production reaching lower atmospheric levels causes the depletion of many species.

\begin{figure*}[h!]
\centering
\includegraphics[scale=0.35]{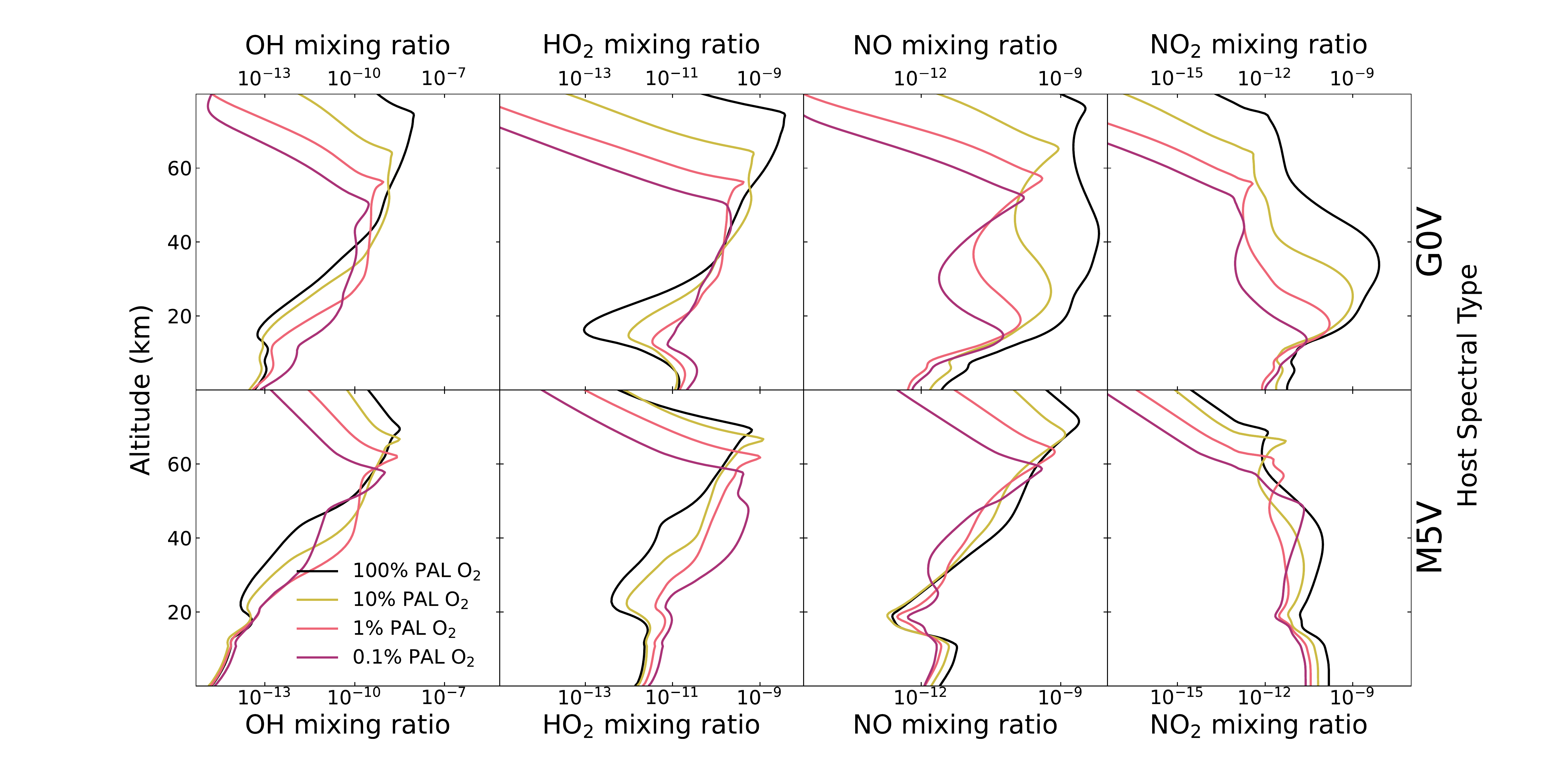}
\caption{Mixing ratio profiles of HO$_x$ (OH + HO$_2$) and NO$_x$ (NO + NO$_2$) species that power the catalytic cycles which destroy \oz\ for models with 100\%, 10\%, 1\%, and 0.1\% PAL \om.  Here we show models for the hottest (G0V) and coolest (M5V) stellar hosts.  For decreasing \om\ levels photolysis reaches lower levels of the atmosphere,  causing OH and HO$_2$ formation to occur at lower levels.  The HO$_x$ is not significantly impacted by these changes since \oz\ formation is pushed deeper into the atmosphere as well.  NO$_x$ species experience depletion via N$_2$O and NO photolysis, although are less affected for models around cooler stars with lower photolysis rates.  The efficiency of the NO$_x$ catalytic cycle decreases consistently for all host stars with decreasing \om\ levels.  See Sect.~\ref{sec:catalyticresults} for full details. \label{fig:catalytic}}
\end{figure*}

H$_2$O is increasingly depleted for decreasing levels of \om\ due to both photolysis in the atmosphere and \od\ reactions lower in the atmosphere.  Both of these reactions create the OH radical while removing H$_2$O,
\be
\m{H}_2\m{O} + \m{h}\nu \rightarrow \m{H} + \m{OH},
\label{r:PH2O}
\ee
\vspace{-0.7cm}
\begin{equation}
\m{H}_2\m{O} + \m{O(}^1\m{D)} \rightarrow \m{OH} + \m{OH}.
\tag{\ref{r:H2O_OH}}
\end{equation}
On modern Earth, Reaction~\ref{r:H2O_OH} is the primary source of OH in the stratosphere, which is a major sink for several species.  As H$_2$O levels in the upper atmosphere drop with decreasing \om, this causes OH production to move to lower levels of the atmospheres as seen in Fig.~\ref{fig:catalytic}.  Upper atmospheric depletion of H$_2$O and OH production at lower altitudes is seen more strongly for hotter host stars, as they have higher incident UV for photolysis and \od\ creation.

CH$_4$, an important biosignature gas, is also depleted in the upper atmosphere for models around all host stars, primarily though oxidation via OH (created via reactions with H$_2$O), along with photolysis and reactions with \od,
\be
\m{CH}_4 + \m{OH} \rightarrow \m{CH}_3 + \m{H}_2\m{O},
\label{r:CH4_H2O}
\ee
\vspace{-0.7cm}
\be
\m{CH}_4 + \m{O(}^1\m{D)} \rightarrow \m{CH}_3 + \m{OH}.
\label{r:CH4_OH}
\ee
In the upper atmosphere, depletion is dominated by photolysis.   Reaction~\ref{r:CH4_H2O} is both the main sink of stratospheric CH$_4$ and OH depletion on modern Earth, with CH$_4$ and OH acting as a major sinks for each other.  Reactions with \od\ and OH occur deeper in the atmosphere for decreasing \om\ levels as both these radicals are produced at lower altitudes.  Reaction~\ref{r:CH4_OH} is an additional source of OH in the lower stratosphere/troposphere.  CH$_4$ depletion is limited to the upper stratosphere for model atmospheres around cooler hosts, although can reach the lower stratosphere for model atmospheres with hotter hosts.

N$_2$O, another potential biosignature gas, experiences extreme depletion for hotter hosts down to the troposphere for lower levels of \om, and significantly less depletion constrained to the upper atmosphere around cooler stars.  This is due primarily to photolysis (Reaction~\ref{r:PN2O_O1D}) in the upper atmosphere, although there are contributions from interactions with \od\ as well,
\begin{equation}
\m{N}_2\m{O} + \m{O(}^1\m{D)}    \rightarrow    \m{NO}    +  \m{NO}.
\tag{\ref{r:N2O_NO}}
\end{equation}
Depletion rates of N$_2$O vary significantly between different stellar hosts due to the strong dependence of incident UV flux on N$_2$O destruction.

In summary, the majority of changes to an atmosphere as \om\ decreases are caused by increased photolysis rates as \om\ UV shielding decreases as well as \od\ production from either \om, \oz, N$_2$O, or CO$_2$ photolysis occurring at lower levels of the atmosphere.  These effects cause upper atmospheric depletion of H$_2$O, CH$_4$, and N$_2$O, with more depletion for hotter stellar hosts with stronger UV fluxes.

\subsection{Impact of varying \om\ on catalytic cycles \label{sec:catalyticresults}}

Varying \om\ levels impacts HO$_x$ (OH + HO$_2$) and NO$_x$ (NO + NO$_2$) species, which are the main contributors of catalytic cycles that destroy \oz\ (see Sect.~\ref{sec:cc} for details).  Mixing ratio profiles of these species are shown in Fig.~\ref{fig:catalytic} for our hottest and coolest host stars.  Once again, the impact on these species as \om\ levels are decreased is controled by photolysis reaching deeper levels of the atmosphere along with \od\ production moving to lower levels as well.

As \om\ decreases, HO$_x$ species (OH and HO$_2$) in all models decrease in the upper atmosphere, but increase in the lower atmosphere.  OH production via reactions with H$_2$O (Reactions~\ref{r:H2O_OH}, \ref{r:PH2O}) and CH$_4$ (Reaction~\ref{r:CH4_OH}) occur at lower altitudes for lower \om\ levels, especially since both H$_2$O and CH$_4$ are depleted in the upper atmosphere from photolysis.  This ``pushing down'' of HO$_x$ species is more noticeable for hotter host stars with higher photolysis rates.  Also note that stars with lower FUV/NUV ratios can better remove \oz\ via the HO$_x$ catalytic cycle, as FUV wavelengths create \oz, while NUV wavelengths photolyze H$_2$O to form OH.  However, for all host stars the efficiency of the HO$_x$ catalytic cycle of \oz\ destruction is not largely impacted for different \om\ and \oz\ abundances since OH and HO$_2$ move down in the atmosphere along with \oz\ concentrations.  Decreased \om\ UV shielding and increased photolysis does not destroy HO$_x$ species, but rather converts them into other HO$_x$ species.  When HO$_2$ is photolyzed,
\be
\m{HO}_2 + \m{h}\nu \rightarrow \m{OH} + \m{O},
\label{r:PHO2}
\ee
it creates OH.  The OH radical itself is extremely reactive with a short lifetime, and typically will react quickly with other species or react with \oz\ to create HO$_2$ (Reaction~\ref{r:HOx_OH}).

Although with decreasing \om\ abundance HO$_x$ species are formed lower in the atmosphere rather than destroyed by photolysis, NO$_x$ species (NO + NO$_2$) can be depleted via photolysis.  The main source of NO$_x$ in the stratosphere is via N$_2$O reactions with \od.  However, as shown in Fig.~\ref{fig:genchem}, N$_2$O is significantly depleted in the atmosphere via photolysis, especially for hotter host stars.  While H$_2$O, the primary source of HO$_x$ species in the stratosphere creates HO$_x$ during photolysis, N$_2$O, the primary source of NO$_x$ species, does not.  Instead it creates N$_2$ and \od\ (Reaction~\ref{r:PN2O_O1D}), cutting off the main source of NO production from N$_2$O.  As for NO$_x$ species themselves, NO$_2$ photolysis creates more NO, while NO photolysis simply breaks the molecule apart,
\be
\m{NO}_2 + \m{h}\nu \rightarrow \m{NO} + \m{O},
\label{r:PNO2}
\ee
\vspace{-0.7cm}
\be
\m{NO} + \m{h}\nu \rightarrow \m{N} + \m{O} ,
\label{r:PNO}
\ee
causing NO photolysis to be a sink of NO$_x$.  NO can be formed once again via reactions between N atoms and \om\ molecules,
\be
\m{N} +\m{O}_2 \rightarrow \m{NO} + \m{O},
\label{r:N_NO}
\ee
although the rate of NO photolysis is faster than Reaction~\ref{r:N_NO}, causing it to be a gradual sink of NO$_x$.  Often the N atom created by Reaction~\ref{r:PNO} will remove NO$_x$ via,
\be
\m{NO} +\m{N} \rightarrow \m{N}_2 + \m{O},
\label{r:NO_N}
\ee
or the N atoms will recombine with other N atoms,
\be
\m{N} +\m{N} \rightarrow \m{N}_2 ,
\label{r:N}
\ee
with this reaction becoming more efficient as \om\ levels drop. This sink via NO photolysis has less of an impact on cooler host stars with lower photolysis rates, hence less NO$_x$ depletion. 

As seen in Fig.~\ref{fig:catalytic}, NO$_x$ species are depleted throughout the atmosphere for the G0V host star, while  the M5V host star experiences less NO$_x$ depletion, and actually an increase in NO in the lower stratosphere.  This is due to primarily to lower photolysis rates for the cooler M5V star which depletes less N$_2$O and NO (Reaction~\ref{r:N2O_NO}).  However, for model atmospheres around all host stars the ability of the NO$_x$ catalytic cycle to deplete \oz\ diminishes consistently with decreasing \om\ levels, even for cooler stellar host models with less NO$_x$ depletion.

In summary, the HO$_x$ catalytic cycles are not hugely impacted by decreasing \om\ because increased photolysis rates tend to push HO$_x$ species to lower altitudes rather than destroy them. However, NO$_x$ catalytic cycles decrease in efficiency with lower \om\ levels since photolysis of N$_2$O and NO remove NO$_x$ from the atmosphere.

{\singlespace
\begin{table*}[h!]
\centering
\footnotesize
\caption{UV Integrated Fluxes \label{tab:UV_all}}
\begin{tabular}{crrrr|rrr|rrr}
Spectral & O$_2$ MR & \multicolumn{3}{c}{UVA 315 - 400 nm (W/m$^2$)}  & \multicolumn{3}{c}{UVB 280 - 315 nm (W/m$^2$)} & \multicolumn{3}{c}{UVC 121.6 - 280 nm (W/m$^2$)}\\
\cline{3-5} \cline{6-11}
Type & (\% PAL) & TOA & Surface & \% to surf. & TOA & Surface &\% to surf. & TOA & Surface &\% to surf.  \\
\hline
\hline
G0V 	 & 100	 & 96.6	 & 77.9	 & 80.7	 & 22.4	 & 1.5	 & 6.7	 & 11.2	 & 3.8e-27	 & 3.4e-26 \\
G0V 	 & 10	 & 96.6	 & 76.7	 & 79.4	 & 22.4	 & 1.4	 & 6.2	 & 11.2	 & 1.8e-08	 & 1.6e-07 \\
G0V 	 & 1	 & 96.6	 & 77.5	 & 80.3	 & 22.4	 & 2.4	 & 10.7	 & 11.2	 & 1.7e-04	 & 1.5e-03 \\
G0V 	 & 0.1	 & 96.6	 & 78.6	 & 81.4	 & 22.4	 & 5.9	 & 26.3	 & 11.2	 & 1.7e-02	 & 1.5e-01 \\
\hline
Sun 	 & 100	 & 82.9	 & 67.5	 & 81.4	 & 16.2	 & 1.6	 & 10.2	 & 6.7	 & 2.8e-21	 & 4.1e-20 \\
Sun 	 & 10	 & 82.9	 & 66.5	 & 80.2	 & 16.2	 & 1.6	 & 9.6	 & 6.7	 & 3.3e-08	 & 5.0e-07 \\
Sun 	 & 1	 & 82.9	 & 67.1	 & 81.0	 & 16.2	 & 2.7	 & 16.4	 & 6.7	 & 2.3e-04	 & 3.5e-03 \\
Sun 	 & 0.1	 & 82.9	 & 67.7	 & 81.7	 & 16.2	 & 5.6	 & 34.8	 & 6.7	 & 1.5e-02	 & 2.3e-01 \\
\hline
K2V 	 & 100	 & 34.2	 & 28.0	 & 81.9	 & 4.8	 & 0.68	 & 14.1	 & 1.4	 & 1.1e-18	 & 8.0e-17 \\
K2V 	 & 10	 & 34.2	 & 27.6	 & 80.8	 & 4.8	 & 0.74	 & 15.4	 & 1.4	 & 1.8e-08	 & 1.3e-06 \\
K2V 	 & 1	 & 34.2	 & 27.8	 & 81.4	 & 4.8	 & 1.2	 & 25.4	 & 1.4	 & 1.0e-04	 & 7.3e-03 \\
K2V 	 & 0.1	 & 34.2	 & 28.0	 & 81.8	 & 4.8	 & 2.2	 & 44.8	 & 1.4	 & 1.0e-02	 & 7.2e-01 \\
\hline
K5V 	 & 100	 & 15.3	 & 12.8	 & 83.6	 & 0.68	 & 0.10	 & 14.4	 & 0.16	 & 2.1e-21	 & 1.3e-18 \\
K5V 	 & 10	 & 15.3	 & 12.6	 & 82.6	 & 0.68	 & 0.14	 & 20.1	 & 0.16	 & 8.4e-09	 & 5.1e-06 \\
K5V 	 & 1	 & 15.3	 & 12.7	 & 82.9	 & 0.68	 & 0.23	 & 33.6	 & 0.16	 & 5.4e-05	 & 3.3e-02 \\
K5V 	 & 0.1	 & 15.3	 & 12.7	 & 83.0	 & 0.68	 & 0.34	 & 50.4	 & 0.16	 & 4.7e-03	 & 2.8e+00 \\
\hline
M5V 	 & 100	 & 1.6	 & 1.3	 & 83.8	 & 3.5e-02	 & 6.5e-03	 & 18.4	 & 2.7e-02	 & 9.8e-21	 & 3.6e-17 \\
M5V 	 & 10	 & 1.6	 & 1.3	 & 83.1	 & 3.5e-02	 & 1.0e-02	 & 28.8	 & 2.7e-02	 & 4.1e-09	 & 1.5e-05 \\
M5V 	 & 1	 & 1.6	 & 1.3	 & 83.4	 & 3.5e-02	 & 1.6e-02	 & 45.7	 & 2.7e-02	 & 3.8e-05	 & 1.4e-01 \\
M5V 	 & 0	 & 1.6	 & 1.3	 & 83.4	 & 3.5e-02	 & 2.0e-02	 & 56.3	 & 2.7e-02	 & 1.1e-03	 & 4.2e+00 \\
\hline
\hline
\end{tabular}
\vspace{-0.2cm}
\tablefoot{
Abbreviations: MR = mixing ratio; PAL = present atmospheric level; TOA =  top of atmosphere}
\end{table*}
}

\subsection{Surface UV flux for different \om\ and \oz\ levels \label{sec:UV}}

\begin{figure*}[h!]
\centering
\includegraphics[scale=0.48]{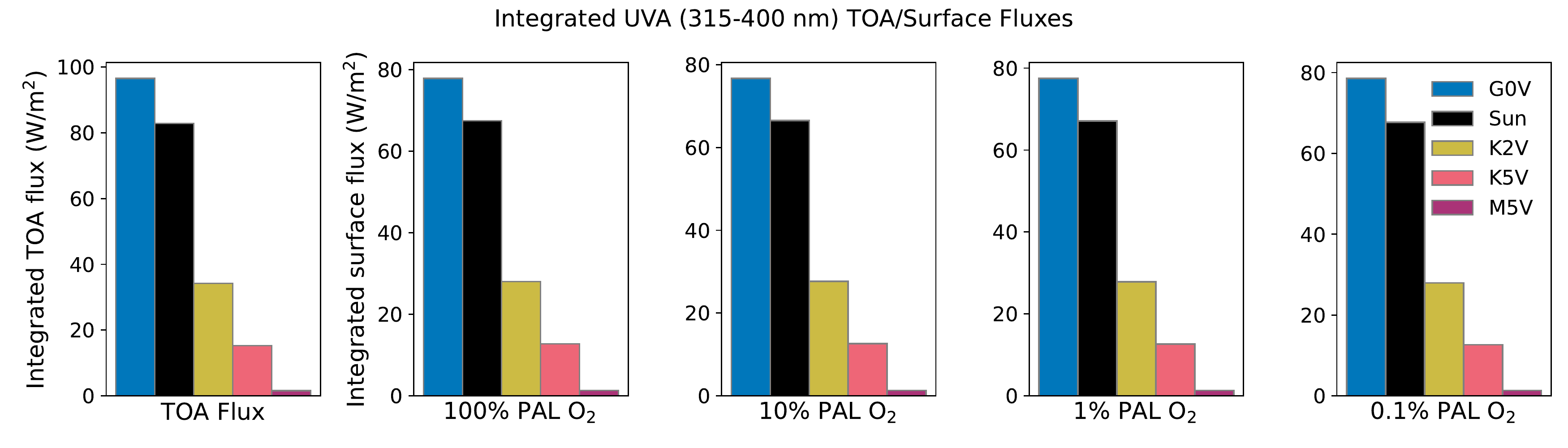}\\
\includegraphics[scale=0.48]{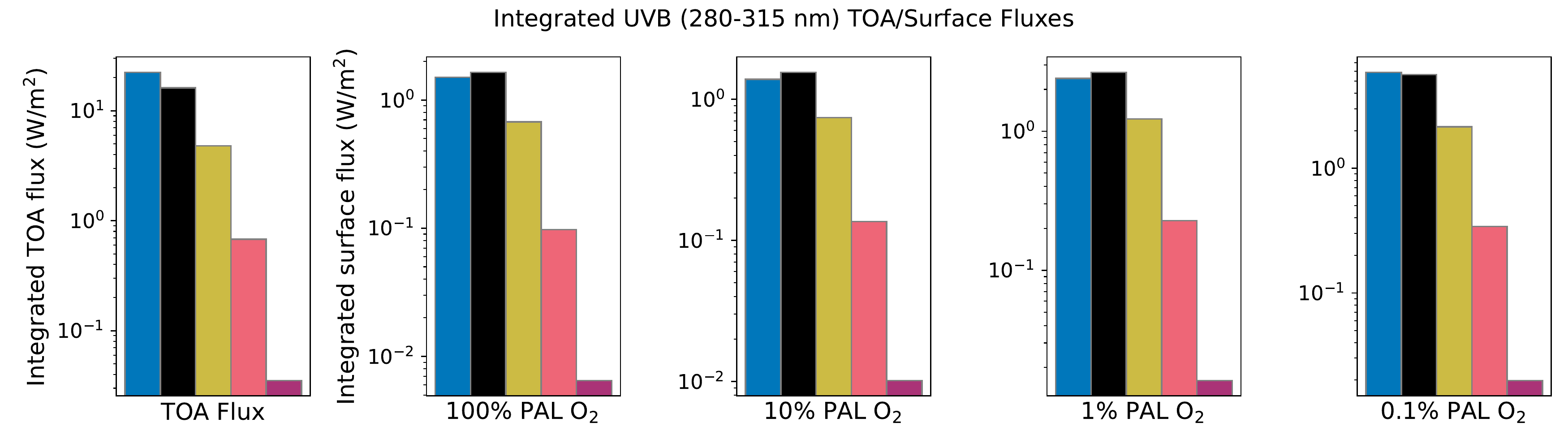}\\
\includegraphics[scale=0.48]{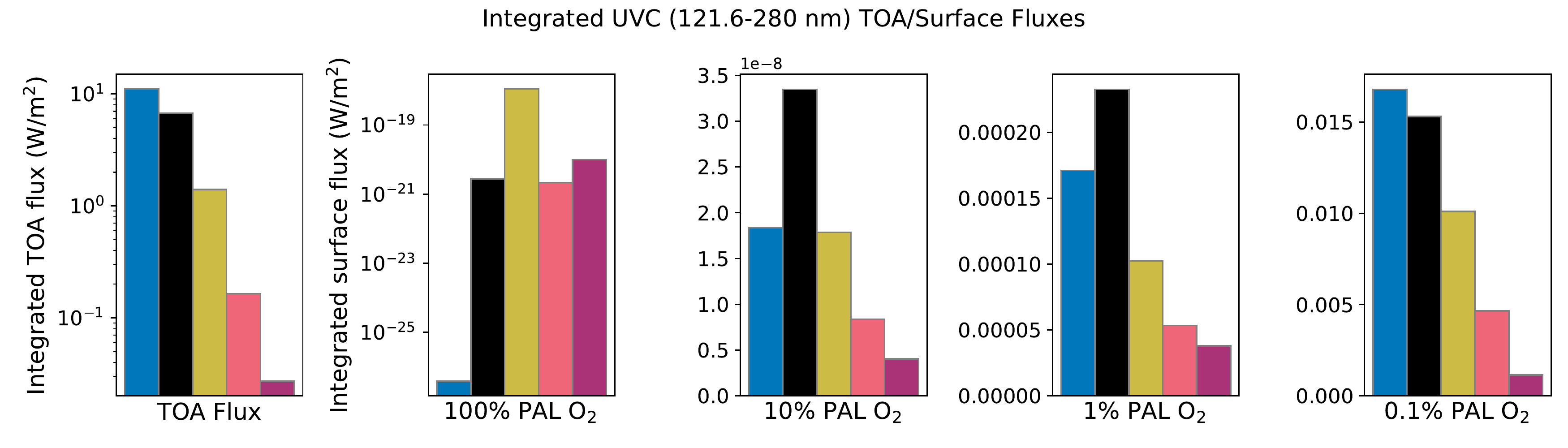}
\caption{Top-of-the atmosphere (TOA) and surface fluxes for UVA (top), UVB (middle), and UVC (bottom) wavelengths for 100\%, 10\%, 1\%, and 0.1\% PAL \om\ model atmospheres for all stellar hosts. We note the differences in y-axis scale for different subplots.  UVA flux is only slightly shielded by \oz\ so surface fluxes scale roughly with TOA fluxes.  UVB flux is partially shielded by \oz\ and therefore allows the G0V-hosted models to receive less surface UVB than Sun-hosted models despite higher TOA UVB due to more efficient \oz\ production.  UVC surface fluxes for high \om\ levels are strongly influenced by \oz\ abundance due to strong  UV shielding from \oz\ in this wavelength range.  For models with only 0.1\% PAL \om\ levels all surface UV fluxes begin to converge to TOA values as the shielding of \oz\ is significantly decreased.  See Sect.~\ref{sec:UV} for full details.
\label{fig:surfUV}}
\end{figure*}

\texttt{Atmos} was used to calculate the amount of UV flux reaching the planetary surface in each model atmosphere.  Surface UV flux is strongly dependent on incident stellar UV flux and the amount of UV shielding from both \om\ and \oz.  High UV fluxes can cause substantial damage to biological organisms, hence UV surface environments will be critical for determining surface habitability.  These results are summarized in Table~\ref{tab:UV_all} and shown in Fig.~\ref{fig:surfUV}.  Surface UV fluxes calculated here using a zenith angle of 60$^\circ$ (see Sect.~\ref{sec:atmos}).

Integrated surface UV fluxes are broken up into three biologically relevant wavelength regimes: UVA, UVB, and UVC.  UVA flux (315-400 nm)  is the lowest energy type of UV and is only partially shielded by \oz, so a large percentage of incident UVA on modern Earth reaches the planetary surface.  UVB (280-315 nm) is more harmful for life, contributing to sun burn and skin cancer in humans and damage to other organisms (e.g., \citealt{kies01}).  UVB is shielded much more efficiently by \oz\ than UVA, with a smaller fraction of incident UVB reaching the surface of modern Earth.  UVC (121.6-280 nm) is capable of causing DNA damage, but is fortunately shielded almost entirely by \oz\ on modern Earth.  Ozone is most efficient at shielding UV in this wavelength region, as evidenced by the \oz\ absorption cross sections shown in Fig.~\ref{fig:XS}.  We note that \om\ photolysis, the first step in \oz\ formation (Reactions~\ref{r:PO2_O}, \ref{r:PO2_O1D}), requires a UVC photon ($\lambda <$ 240 nm), allowing \om\ to contribute partially to UVC shielding.  However, since \oz\ is the primary shielder of UVC, the requirement of a UVC photon to produce \oz\ creates interesting correlations between incident and surface UVC flux.

Because UVA is not strongly shielded by \oz, UVA surface fluxes for all models are closely correlated with the amount of incident UVA flux (see Table~\ref{tab:UV_all} and Fig.~\ref{fig:surfUV}). For all host stars at \om\ levels of 100\%, 10\%, 1\%, and 0.1\%  PAL the amount of incident UVA that reaches the surface of these model planets is roughly $\sim$80\% for all cases.  Because \oz\ plays only a small role in UVA shielding, all model results are quite similar.

UVB surface fluxes are significantly more variable because \oz\ shielding is much more important for these wavelengths. Although the G0V host star provides a higher incident UVB flux than the Sun, G0V-hosted models still maintain slightly less surface UVB flux until \om\ decreases to 0.1\% PAL, at which point the G0V and Sun model surface fluxes become roughly equal. This is due to the larger amount of \oz\ created by the G0V host star compared to the Sun, which allows for stronger UVB shielding (see Fig.~\ref{fig:O3-O2_MR_all} for \om-\oz\ relationship).  The percentage of incident UVB flux that reaches the planetary surface varies significantly between different host stars.  For our hottest host star (G0V) the amount of incident UVB flux reaching the planetary surface increases from 6.7\% to 26.3\% as \om\ levels drop from 100\% to 0.1\% PAL.  As a result of less \oz\ shielding, these percentages are higher for our coolest host star (M5V), which experiences an increase of 18.4\%  to 56.3\% of incident UVB reaching the surface as \om\ decreases from 100\% to 0.1\% PAL. Even though the cooler stellar hosts allow a higher percentage of UVB flux to travel through the atmosphere, they still maintain lower surface UVB values than hotter hosts due to their weaker incident UVB flux.

The strong reliance of UVC absorption on \oz\ abundance, along with the fact that \oz\ creation requires UVC photons, leads to some unexpected UVC surface flux results.  A striking consequence of this is that while the G0V host star provides the highest incident UVC flux of all our host stars, it maintains the lowest surface UVC flux for the  100\% PAL \om\ model by several orders of magnitude, while the much cooler K2V stellar host model experiences the highest surface UVC flux.  The much higher incident UVC flux of the G0V host causes much faster \oz\ production than other host stars, allowing for UVC shielding strong enough to counteract the high incident UVC flux.  Another interesting result for the 100\% PAL \om\ case is that the M5V model has a slightly higher UVC surface flux than the K5V model, despite the fact that the M5V model has the lowest incident UVC flux.  Again, this effect is due to the higher \oz\ abundance of the K5V-hosted planet, created by the stronger incident UVC flux.  For all host stars, the atmospheric models with 100\% PAL \om\ allowed only extremely tiny fractions of incident UVC flux reach the surface ($<10^{-17}$\% in all cases).

UVC surface fluxes for models with 10\% and 1\% PAL \om\ have similar trends when comparing stellar hosts.  Sun-hosted models had the largest surface UVC fluxes in both scenarios.  Notice that although the model atmospheres hosted by the G0V star and the Sun have higher \oz\ levels for the 10\% PAL \om\ cases compared to their 100\% PAL \om\ cases, overall UVC shielding is significantly less for the 10\% PAL \om\ cases due to the lesser contribution of \om\ absorbing
photons  with wavelengths less than 240 nm.  Even though the G0V host and the Sun produce much larger amounts of \oz\ than cooler stars, for low \om\ levels the combined decrease in \om\ and \oz\ UV shielding causes them to have higher surface UVC fluxes than cooler stars that produce significantly less \oz.  For model atmospheres with  \om\ levels of only 0.1\% PAL,  surface UVC levels begin to converge to the incident UVC flux as \om\ and \oz\ levels have dropped enough that they shield UVC far less effectively.  It has previously been suggested that the usefulness in the ability of \oz\ to shield UV drops off drastically at these \om\ values (e.g.,\ \citealt{segu03}).  However, due to CO$_2$ shielding, all models in this study had virtually no photons with wavelengths less than 200 nm reach the planetary surface, even with the lowest \om\ abundance modeled (0.01\% PAL).

Although the model atmospheres hosted by the hottest stars create the highest levels of \oz, they constantly experience the highest  UVA and UVB surface fluxes due to the limited shielding abilities of \oz\ in these wavelength ranges.  However, for the far more damaging UVC wavelengths at 100\% PAL \om\ it is the G0V host star that provides the lowest UVC surface flux by orders of magnitude, with the Sun-hosted models having comparable UVC surface flux to cooler host star models. Somewhat ironically, for lower \om\ levels of 1 to 10\% PAL, it is the Sun that is the host star with the least ``hospitable'' conditions for surface life with the highest UVC surface fluxes.  As \om\ drops to 0.1\% PAL  UVC surface flux will begin to converge to the incident UVC flux as \om\ and \oz\ shielding drops dramatically.  However, though life on modern Earth requires a substantial \oz\ layer for UV protection, it is important to remember that evidence for life on Earth dates back to 3.7 Gyr ago \citep{rosi99}, long before the \om\ levels rose during the Great Oxidation Event 2.5 Gyr ago.  The lack of significant atmospheric UV shielding may prevent life as we know it, but it does not rule out its existence. Life could exist, for instance, underwater, at a depth in which significant damaging UV has been absorbed by water (e.g., \citealt{cock07}).

\begin{figure*}[h!]
\begin{center}
\includegraphics[scale=0.45]{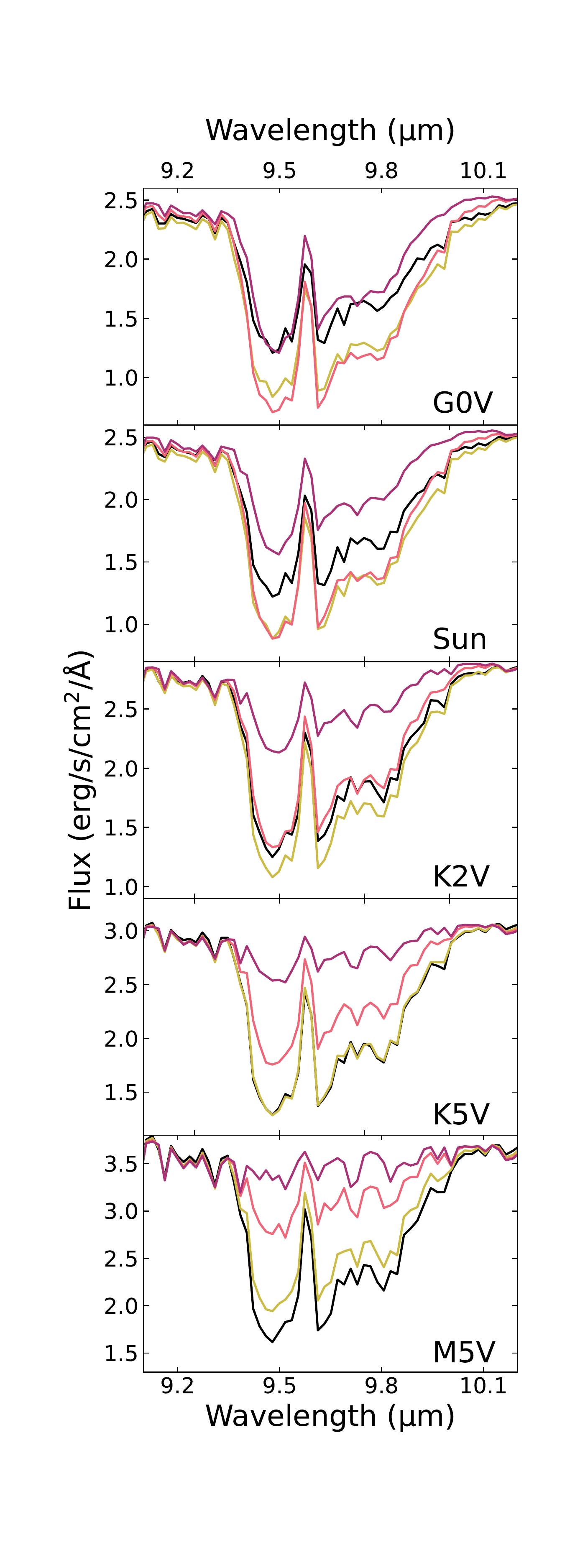}
\hspace{-0.5cm}
\includegraphics[scale=0.45]{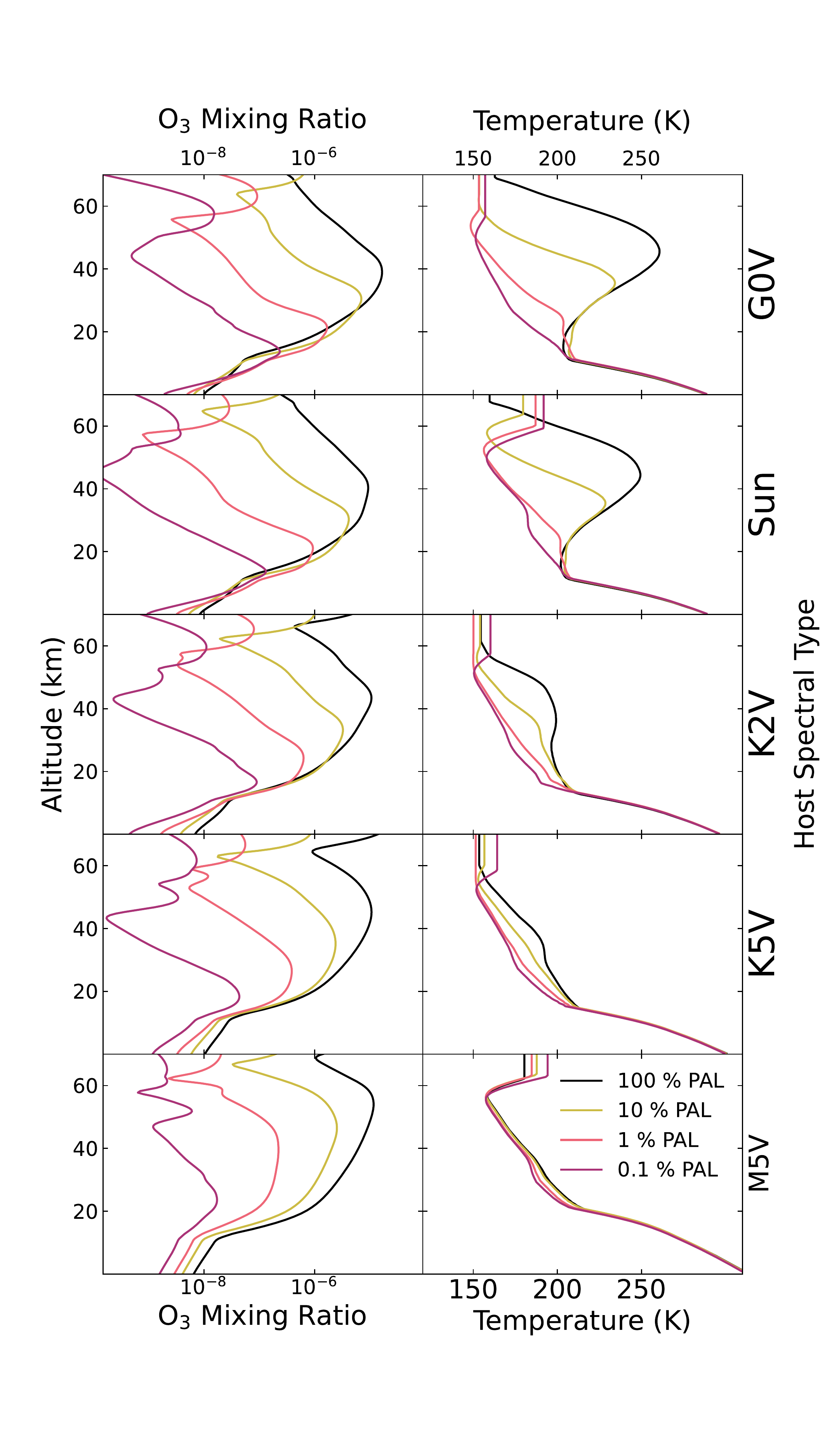}
\end{center}
\caption{Emission spectra, \oz\ mixing ratio profiles, and temperature profiles for model atmospheres with 100\%, 10\%, 1\%, and 0.1\% PAL \om\ for all host stars.  The depth of spectral features in emission spectra is dependent on the temperature difference between the absorbing and emitting layers of the atmosphere, causing the  \oz\ feature depth to strongly correlate  with the temperature difference between the stratosphere and planetary surface.  However, since \oz\ is responsible for the majority of stratospheric heating, this results in higher \oz\ abundances (with more stratospheric heating) having shallower spectral features than atmospheres with less \oz\ and cooler stratospheres.  For full details see Sect.~\ref{sec:emissionresults}. We note that temperatures for parts of the atmosphere above 1 mbar are held constant as that is the maximum height computed by the climate code (see Sect.~\ref{sec:atmos} for more details). \label{fig:emission}}
\end{figure*}

\subsection{\oz\ spectral features for different \om\ levels \label{sec:emissionresults}}

Emission spectra from our model atmospheres are shown in Fig.~\ref{fig:emission}, zoomed in on the  primary MIR \oz\  feature at 9.7 $\mu$m, along with the corresponding \oz\ mixing ratio and temperature profiles, which are necessary for interpreting the features.  The temperature difference between the absorbing and emitting layers of the planet's atmosphere, rather than the overall abundance of that gaseous species, determines the depth of planetary emission spectrum features. Because \oz\ is a main contributor of stratospheric heating, the strength of \oz\  features has a highly nonlinear relationship to \oz\ abundance.  Once again, we see counterintuitive trends for hotter host stars (G0V, Sun, K2V), and different, more straightforward trends, for cooler host stars (K5V, M5V).

For all host stars, the 0.1\% PAL \om\ case has the shallowest \oz\  feature in emission spectra, but the \om\ level for the deepest feature depends on the host star.  For the G0V-hosted models the \oz\ feature for the 100\% PAL \om\ case has a similar depth to the 0.1\% PAL \om\ case, despite the fact that they have significantly different integrated \oz\ column densities (7.06$\times10^{18}$ cm$^{-2}$ for 100\% PAL \om; 1.23$\times10^{18}$ cm$^{-2}$ for 0.1\% PAL \om).  For the two hottest host stars (G0V, Sun), the 10\% and 1\% PAL \om\ cases are the deepest features.  The strong features of the 10\% PAL \om\ models are not surprising since both the G0V and Sun models have higher \oz\ abundances at  10\% PAL  than at 100\% PAL \om, but the 1\% PAL \om\ models have significantly less \oz\ than both the 10\% and 100\% PAL \om\ cases (see Fig.~\ref{fig:O3-O2_MR_all} for reference).  Conversely, \oz\ features for the coolest host star models correspond more intuitively to \om\ and \oz\ levels, with the highest \om\ and \oz\ abundances having the deepest features, and the lowest \om\ and \oz\ abundances having the shallowest features.

The relationship between the depth of \oz\ spectral features and actual \oz\ abundance is dictated by atmospheric temperature profiles.  The temperature difference between the emitting and absorbing layers of a gaseous species determines feature depth, therefore \oz\ feature depth is determined by the temperature difference between the altitude of peak \oz\ concentration in the stratosphere and the planet's surface temperature.  Because  \oz\ NUV absorption is a dominant source of stratospheric heating, a higher \oz\ concentration with significant incident NUV flux for \oz\ to absorb results in higher stratospheric temperatures and thus a shallower spectral feature.  This explains why an atmosphere with a large amount of \oz\ and high incident NUV flux (and more stratospheric heating) has a weaker \oz\  feature than an atmosphere with less \oz\ and weaker incident NUV, but a larger temperature difference between the stratospheric and surface temperatures.  Cooler host star models with less \oz\ formation and lower incident NUV flux have significantly less stratospheric heating (Fig.~\ref{fig:emission}), and therefore \oz\ spectral feature depths which correspond more strongly with the actual abundance of \oz\ in their atmospheres.  

In summary,  in order to interpret \oz\  features in planetary emission spectra and retrieve the \oz\ (and \om) abundances it will require modeling of the atmospheric temperature profiles.  Both photochemistry and climate modeling will be essential in this process.

\section{Discussion \label{sec:discussion}}

\subsection{Comparison to other studies}

Multiple studies have explored \oz\ formation in Earth-like atmospheres using a variety of models, each providing valuable insight on the \om-\oz\ relationship.  Here, we briefly describe relevant past studies on this topic.  With 1D modeling of Earth's atmosphere, early \oz\ studies revealed the nonlinear link between \oz\ and \om.  Both \cite{ratn72} and \cite{levi79} discuss the phenomenon of the \oz\ layer moving down in the atmosphere as \om\ levels decreased (see Sect.~\ref{sec:chemresults} for details on this process) and agreed on peak \oz\ abundance occurring at $\sim$10\% PAL \om. Total \oz\ abundances calculated for these studies differed because they each included different chemical reactions. The model in \cite{ratn72} contained only the Chapman mechanism, while \cite{levi79} additionally HO$_x$ and NO$_x$ catalytic cycle destruction of \oz\ in their model.  Later \cite{kast85} replicated the \oz\ peak in abundance at lower \om\ levels using a more sophisticated model including chemistry beyond the Chapman mechanism and catalytic cycles, incorporating 20 gaseous species overall.  They predicted maximum \oz\ production to occur at 50\% PAL \om, a higher \om\ estimate than previously.  It is important to note that none of these studies included a climate model to calculate self-consistent atmospheric temperatures.  Because the Chapman mechanism is temperature dependent, this helps account for discrepancies with later \oz\ calculations.

Studies in later years began to model \oz\ production in planetary atmospheres with different types of host stars.  \cite{segu03} used what they described as a ``loosely coupled'' 1D climate and photochemistry code (partially based off the \citealt{kast85} model) for different \om\ levels around F2V, G2V, and K2V type stars.  We note that this model is a predecessor of the model used in this study: \texttt{Atmos}. No host star displayed a peak \oz\ abundance at an \om\ level less than 100\% PAL, but this is likely because the modeled \om\ levels were evenly spaced on a logarithmic scale from 0.001-100\% PAL \om, whereas more finely spaced \om\ levels are required to capture this effect.  Atmospheric chemistry and temperature profiles computed in \cite{segu03} are similar to our Sun and K2V host star models.  There are slight differences in total \oz\ abundance in these models compared to those in this study (our models tend to have lower \oz), although this is likely due to differences in input UV spectra, boundary conditions, and model updates.  Overall this is the most similar study to ours in terms of variety of \om\ levels and host stars.

Other studies have also modeled \oz\ formation in Earth-like planetary atmospheres around different stellar hosts.  The effect of varying orbital separations inside the habitable zone on \oz\ formation was explored for F2V, G2V, and K2V hosts (same as \citealt{segu03}) in \cite{gren07}, and for a variety of M dwarfs in \cite{gren14}.  Both these studies used the 1D ``loosely coupled'' climate and photochemistry model developed in \cite{segu03}.  An increase in star-planet separation for FGK stars caused cooler atmospheric temperatures, which correlated to an increase in \oz. This is because the 3-body reaction that creates \oz\ (Reaction~\ref{r:O2M}) is faster at cooler temperatures. However, this \oz\ increase was not large because larger orbital distances also caused higher levels of HO$_x$ and NO$_x$ species which destroy \oz\ \citep{gren07}.  When repeating this study for M dwarfs  they found what they described as a  ``Goldilocks'' effect in which there was a range of UV that was best for creating the most detectable \oz.  If incident  UV flux is too low it will create small amounts of  \oz\ making it harder to detect, but if the UV flux is too high it will create enough \oz\ to cause significant stratospheric heating, making it more difficult to detect in planetary emission spectra (see Sect.~\ref{sec:emissionresults} details on this phenomenon).  M7V spectral types were found to produce the amount of UV that was ``just right'' in creating detectable amounts of \oz\ \citep{gren14}.  Although these models were run only at 100\% PAL \om, their results are consistent with this study.

The impact of stellar host UV on \oz\ formation has also been modeled  using \emph{Exo-Prime}, a 1D coupled climate and photochemistry originally based off the same codes as \texttt{Atmos}, for Earth-like planets orbiting FGKM stars \citep{rugh13}, M dwarfs \citep{rugh15}, cool white dwarfs \citep{koza18}, and red giants \citep{koza19}.  However, all these studies were constrained to \om\ abundances of 100\% PAL, although our corresponding models results are consistent.  Another \emph{Exo-Prime} study \citep{rugh15b} modeled Earth at different points throughout geological history for FGKM stars, including four different \om\ abundances, although the large variations in abundances of many gaseous species (i.e., CH$_4$, CO$_2$) does not allow for a straightforward comparisons of results with our study.

Of the 1D models discussed in this study, only \texttt{Atmos} models photochemistry in the mesosphere and lower thermosphere (up to an altitude of 100 km), whereas other photochemistry models are limited to atmospheric heights below $\sim$65 km \citep{kast80,kast85,segu03,gren07,gren14,rugh13,rugh15,rugh15b,koza18,koza19}.  This is relevant for \oz\ formation because of the ``secondary \oz\ layer'' on Earth above the stratosphere (see details in Sect.~\ref{sec:chem} and \ref{sec:chemresults}).  High energy photons ($\lambda <$ 175 nm) are normally absorbed above the stratosphere by \om\ photolysis, creating the \od\ radical in the process (Reaction~\ref{r:PO2_O1D}).  This could yield different findings for a photochemistry model that does not include higher altitudes since the high-energy photons will then be absorbed at far lower altitudes than in reality. This would change both the \od\ and \oz\ atmospheric profiles, and account for the  differences in \oz\ production the we see from different models.  However, overall results from the \cite{segu03} model, \emph{Exo-Prime}, and \texttt{Atmos} remain fairly consistent.

Along with 1D models, \oz\ formation has been modeled in 3D.  In reality,  \oz\ formation and abundance is dependent on both the atmospheric latitude and time of day.  On the night side of a planet, \oz\ cannot be generated by the Chapman method nor destroyed by photolysis.  During the day \oz\ is created most efficiently at the equator where incident UV flux is highest, and  then is transported toward the poles by Dobson-Brewer circulation, causing peak \oz\ abundance to vary in altitude depending on the latitude.  On modern Earth, there is only a $\sim$2\% difference in \oz\ between the day and night sides, while planets with differing rotation periods may have more unevenly distributed \oz. Despite the fact that 1D models (like ours) can use a zenith angle to represent the ``average'' of incoming radiation, they cannot accurately predict \oz\ formation and transport for slowly rotating planets, especially ones that are tidally-locked.  Several studies have used 3D modeling to explore \oz\ formation and distribution on tidally locked planets. \cite{proe16} used a 3D climate and photochemistry model to compare \oz\ distribution on modern Earth and a tidally-locked version of Earth.  They found that \oz\ could be distributed to the night-side of the planet and accumulate there in the absence of photolysis.  Hemispheric maps of \oz\ distribution at different phases demonstrated that the amount of detectable \oz\ will be phase-dependent during observations.

Tidally-locked Earth-like planets are most likely found around lower mass stars, where the tidal-locking radius is within the habitable zone, and rotation periods are substantially shorter than the ``tidally-locked Earth around the Sun'' scenario investigated in \cite{proe16}.  \cite{caro18} used a 3D GCM to model how the rapid rotation of a tidally-locked planet would affect \oz\ transport on a planet with a 25-day period.  This faster rotation created  an ``anti-Dobson-Brewer circulation'' effect, with \oz\ accumulating at the equator rather than being transported toward the poles as on modern Earth.  However, this study did not employ a photochemistry model, only circulation effects.

\cite{chen18} and \cite{yate20} also performed 3D modeling of tidally-locked habitable zone planets orbiting M dwarfs, although they incorporated both climate and photochemistry models as well.  \cite{chen18} used CAM-Chem, a 3D model including 97 species in chemistry computations, while \cite{yate20} used the Met Office Unified Model including only Chapman mechanism and HO$_x$ catalytic cycle chemistry.  Both studies found that \oz\ would be transferred to the night-side after being created on the day-side, where it could accumulate to a higher quantity than on the day-side.   HO$_x$ species that were also transported to the night-side would be the primary sink of night-side \oz.  However, \cite{yate20} computed a thinner \oz\ layer  than \cite{chen18}, with the latter's model also computing that \oz\ would exist at higher altitudes.  These differences are likely due to differing chemical networks,  land mass fraction (only \citealt{chen18} had continents), and input stellar spectra.  Overall these results agreed well with \oz\ abundances calculated by \cite{rugh15} for a similar host star, although H$_2$O mixing ratios (important for creating HO$_x$ species) were shown to vary significantly more, showing that 1D models do not include important feedback loops contained within 3D models.

The closest similar 3D modeling work to ours is \cite{cook22}, which uses a 3D climate-chemistry code to model the Earth-Sun system across geological history at various \om\ levels from 0.1\% to 150\% PAL.  Comparing their results to our Sun-hosted models there is agreement between trends in the time-averaged mixing ratios for different gaseous species.  However, a main finding of \cite{cook22} is that their 3D model predicts lower \oz\ abundances for \om\ levels 0.5\%  to 50\% PAL when compared to 1D models, including those calculated here as well as in \cite{segu03} and \cite{rugh15b}.  The cause for these lower estimates from this 3D model is uncertain, although is possibly related to how 1D codes simulate diurnal averages, or different CO$_2$ abundances/boundary conditions.  Further inter-model comparison will be needed in order to clarify these discrepancies \citep{cook22}.

Overall our results for \oz\ formation are consistent with previous 1D studies and roughly similar to time-averaged results from 3D models. Despite this, it is important to remember that the night-sides of slowly rotating and tidally locked planets may have significantly more \oz\ than the day-side, introducing phase-angle dependency on the amount of detectable \oz\ for observations.

\subsection{Factors that impact the \om-\oz\ relationship  \label{sec:o2o3relationship}}

 The \om-\oz\ relationship is strongly dependent on the host star as well as the planetary atmospheric composition.  Here we will briefly describe several ways the \om-\oz\ relationship can diverge from the results in this study. 

We have shown that the \om-\oz\ relationship is highly influenced by the UV spectrum of the host star, both in terms of the total amount of UV flux and the FUV/NUV flux ratio, with FUV primarily responsible for creating \oz, and NUV destroying it.  In this study we selected stellar hosts from a range of spectral types, but have not yet explored the variation of UV activity and FUV/NUV ratios within specific spectral types.  This is of particular importance for K and M dwarfs, as they are subject to larger amounts of UV variability, and thus greater variations in the \om-\oz\ relationship for the planets such stars host (e.g., \citealt{fran13,fran16,youn16,loyd18}).  For instance, the UV spectrum for our M5V host star comes from GJ 876 which displays low amounts of chromospheric activity \citep{fran16}.  If the  stellar host in question was a more active star of a similar type, such as Proxima Centauri (classified as an M5.5V star; \citealt{boya12,angl16}), an orbiting Earth-like planet would be subject to a different \om-\oz\ relationship due to the significant change UV spectral slope of the star.  A more in-depth study of the impact of varying UV activity levels for K and M dwarf planetary hosts will be necessary to fully understand how \oz\ production would vary for different \om\ atmospheric abundances.

Another important aspect of this study to note is that the initial conditions of atmospheric species are kept constant across all models to better understand how the \om-\oz\ relationship differs for different host star spectra.  However, the \om-\oz\ relationship could be altered by a variety of scenarios due to the potentially huge diversity of terrestrial planet atmospheric compositions.

The HO$_x$ and NO$_x$ catalytic cycles are the most prominent sinks for \oz\ on modern Earth, and could significantly impact \oz\ formation if there was an increase or decrease of the species powering these cycles.  Therefore, changes in the amount of stratospheric H$_2$O or N$_2$O would alter the efficiency of \oz\ destruction, as they are the primary sources of stratospheric HO$_x$ and NO$_x$.  On modern Earth H$_2$O is generally prevented from traveling into the stratosphere by the cold trap, although it can be created in the stratosphere via CH$_4$ reactions with OH (Reaction~\ref{r:CH4_H2O}), implying a change in CH$_4$ will additionally impact \oz\ destruction.  The impacts on the \om-\oz\ relationship as these abundances change will be explored at length in the next paper of this series.  Reducing gases in general (e.g., CH$_4$, H$_2$) can impact \om\ and \oz\ levels, whether produced biologically or through volcanic outgassing (e.g., \citealt{hu12,blac14,greg21,cook22}). \oz\ can also be depleted by cometary impacts (e.g., \citealt{marc21}) and through solar flares (e.g., \citealt{pett18}).  In addition, \oz\ can vary throughout different seasons on modern Earth \citep{olso18}.

Oxygen-bearing species in general can also influence the \om-\oz\ relationship, especially in situations where \om\ is produced abiotically via photolysis-driven production (see \citealt{mead17} for full review).  In particular, CO$_2$-rich atmospheres may create significant amounts of \oz\ through CO$_2$ photolysis \citep{hu12,doma14,tian14,harm15,gao15} around host stars with high FUV/NUV flux ratios.   
FUV photons ($\lambda <$ 200 nm) photolyze CO$_2$ ,
\be
\m{CO}_2 + \m{h}\nu \rightarrow \m{CO} + \m{O},
\label{r:PCO2}
\ee
to produce an O atom (or the \od\ radical if $\lambda < 167$ nm; Reaction~\ref{r:PCO2_O1D}).  Oxygen atoms can combine to create \om,
\be
\m{O} + \m{O} + M \rightarrow \m{O}_2 + M,
\label{r:O_O2}
\ee
which can then combine with O atoms to create \oz\ (Reaction~\ref{r:O2M}).  Because \om\ is photolyzed at shorter wavelengths than \oz\ ($\lambda <$ 240 nm, see Fig.~\ref{fig:XS}),  stellar hosts with high incident FUV/NUV flux ratios can allow abiotic \oz\ accumulation without significant corresponding \om\ buildup \citep{hu12,doma14,tian14,harm15}.  In such scenarios the \oz/\om\ ratio would be higher than what would be predicted if \oz\ were formed directly from \om, implying that a high \oz/\om\ ratio could indicate non-biological \oz\ (and \om) creation \citep{doma14}.  However, it remains uncertain which types of stellar hosts would be favorable for this scenario. Some studies find Sun-like stars can accumulate significant \oz\ through CO$_2$ photolysis if outgassing rates of reduced species are low \citep{hu12}, while others restrict this scenario to K and M dwarfs with high FUV/NUV flux ratios \citep{tian14,harm15}.  This scenario might likewise be produced by F star hosts with their strong FUV fluxes, although with enough NUV flux, \oz\ destruction rates could prevent \oz\ buildup \citep{doma14}. Differences in model lower boundary conditions, which control the impact of different \om\ ground sinks, are likely to blame for the disparity in the  capacity of \oz\ to accumulate between different studies.
\citep{doma14,tian14,harm15,mead17}.  Despite uncertainties in \om\ surface sinks, it is clear K/M dwarfs with high FUV/NUV ratios are susceptible, and potentially hotter stars with low abundances of reduced gaseous species.  The effect of a CO$_2$-rich atmosphere on the \om-\oz\ relationship will be highly influenced by the host star and atmospheric abundances of reduced gaseous species.

Another method of creating \oz\ without using the Chapman mechanism is via the  ``smog mechanism'', which can produce \oz\ photochemically using a volatile organic compound (i.e., CH$_4$) and NO$_x$.  This process is responsible for smog pollution often occurring in large cities on modern Earth, but could also have occurred during the Proterozoic (2.5 Ga - 541 Ma) with high levels of CH$_4$ and NO$_x$ \citep{gren06}.  Under some circumstances \cite{gren06} computed that nearly double the amount of \oz\ on modern Earth could have been produced with just 1\% PAL \om\ via the smog mechanism.  Additionally, \cite{gren13} found that the \oz\ smog mechanism may become more efficient than the Chapman mechanism for habitable zone planets around late M dwarf with low UV that is less efficient at \om\ photolysis.  Not only would \oz\ created primarily by the smog mechanism rather than the Chapman mechanism change the \om-\oz\ relationship, but ``smog'' \oz\ can be harmful for life.  Smog mechanism \oz\ is created in the troposphere rather than the stratosphere, and could result in significant ground-level \oz.  Although on Earth our stratospheric \oz\ protects life by shielding harmful UV, ground level \oz\ on a smog-dominated planet can become fatal to Earth organisms at $\sim$1 ppm.

Overall the \om-\oz\ relationship could be subject to large variations based both on the UV spectral slope of the host star, as well as atmospheric composition.  Ozone formation via either CO$_2$ photolysis or hydrocarbon reactions would not be expected to resemble the \om-\oz\ relationship that ``Earth-like'' atmospheres would demonstrate.  However, the FUV/NUV flux ratio of the host star may allow us to rule out certain certain scenarios without observations of the planetary atmosphere.

\subsection{Can we infer \om\ abundance from an \oz\ measurement? \label{sec:O2fromO3}}

Returning to the question that prompted this study: is \oz\ a reliable proxy for \om? Variations in the \om-\oz\ relationship (Sect.~\ref{sec:o2o3relationship}) would increase the difficulty in using \oz\  to infer \om\ abundance, and would require additional atmospheric information to provide the proper context.  For the sake of simplicity, we will discuss the possibility of inferring \om\ from an \oz\ measurement from our ``Earth-like'' models in this paper.  But even in this simplified case where we keep initial conditions of all atmospheric species constant (apart from \om\ and \oz\ and let them adapt to different stellar hosts) precisely determining \om\ from \oz\ is not straightforward due to the nonlinear \om-\oz\ relationship.

Figure~\ref{fig:O3-O2_MR_all} clearly demonstrates that not only does the amount of \oz\ created for different \om\ levels change significantly for different spectral types, but also the trend that the \om-\oz\ relationship will follow as \om\ is changed will depend on the stellar host.  Section~\ref{sec:chemresults} details how planets around hotter stars with higher UV flux (G0V, Sun, K2V) all experience their maximum \oz\ formation efficiency at \om\ levels lower that 100\% PAL, while there is a continuous (albeit nonlinear) decrease of \oz\ production for cooler hosts with lower UV flux (K5V, M5V). Whether a model atmosphere experiences an increase in \oz\ production as \om\ decreases (as seen for hotter stars) is dependent primarily on whether \om\ photolysis ($\lambda <$ 240 nm) can reach deep into the atmosphere. The total amount of \oz\ depends on the FUV/NUV flux ratio of the host star as well, with  FUV flux creating \oz\, while NUV wavelengths will cause its destruction (see Sects.~\ref{sec:stellarspectra} and \ref{sec:chemresults}).  Although the K2V-hosted models demonstrate this effect with the maximum amount of \oz\ production occurring at 55\% PAL \om, for models at \om\ levels near 100\% PAL \om\  the K5V and M5V hosts have higher \oz\ abundances due to their larger FUV/NUV ratios.  Therefore, to predict the \om-\oz\ relationship for a given star even with knowledge of ``Earth-like'' conditions knowing both the total UV emitted and UV spectral slope of the host star will be essential.

Idealized planetary emission spectra of the 9.7 $\mu$m \oz\ features in Fig.~\ref{fig:emission} show a non-trivial relationship between both \om\ and \oz\ abundance and spectral feature depth for hotter stellar hosts,  with more ``straightforward'' correlation of \om\ and \oz\ abundances and feature depth for cooler hosts (K5V, M5V).  This is due to the dependence of feature strength in emission spectra on the atmospheric temperature profile, with \oz\ measurements being particularly complicated by the fact that \oz\ highly influences stratospheric heating.  Measuring \oz\ abundance from an emission spectra will require modeling of atmospheric temperature and pressure profiles for an accurate estimate, especially for stars emitting enough UV capable of creating and absorbing \oz\ for significant stratospheric heating.  Smaller amounts of \oz\ as created by cooler stars have less of an impact on stratospheric heating, and therefore will maintain a more consistent temperature profile even for large variation in \om\ abundance.

Even operating under the unlikely assumption that an accurate measurement of atmospheric \oz\ could be done, inferring \om\ abundance will still not be straightforward, especially for hotter host stars (G0V, Sun, K2V) where \oz\ does not always decrease as \om\ levels decrease.  For example, in our Sun-hosted models the total integrated  \oz\ column density at 150\% PAL \om\ is roughly the same as the amount at 5\% PAL \om\ (4.92$\times10^{18}$cm$^{-2}$ and 4.80$\times10^{18}$cm$^{-2}$, respectively).  This implies that for hotter stellar hosts, it is unlikely \om\ could be well constrained from an \oz\ measurement for relatively high levels of \om.  It could, however, be possible to use an \oz\ measurement to differentiate between pre- and post-GOE \om\ levels, as well as infer the existence of a substantial \oz\ layer providing surface UV shielding.  Due to the consistent decrease in \oz\ abundance with decreasing \om\ for cooler hosts, it appears that inferring \om\ from \oz\ may be much simpler for planets orbiting cooler stars that those around hotter stars.  It is important to note that specific knowledge of the UV spectral slope for cool K and M dwarfs will be extremely important to model \oz\ levels, especially due to the increased likelihood of \oz\ buildup via abiotic means and the diversity of activity levels (hence FUV/NUV flux ratios) around such stars.

\subsection{Is it necessary to constrain \om\ abundance for \oz\ to be a useful biosignature?  \label{sec:O3biosignature}}

Although inferring \om\ levels precisely from \oz\ measurements will not be possible for hot stellar hosts and will still require additional atmospheric context and knowledge of the UV spectral slope for cooler hosts, what does this mean for \oz\ as a biosignature?  Would it be necessary to infer \om\ for \oz\ to be a useful indicator of life, or could it serve as a promising biosignature without precise \om\ information?  Two of the strongest arguments against \om\ as a biosignature are 1) it can be produced abiotically, and 2) it has been at relatively high abundances for only a small fraction of Earth's geological history (see review in \citealt{mead17,mead18}). We  examine these arguments as they pertain to \oz\ as a biosignature.

The multiple proposed pathways for abiotic \om\ production will prevent \oz\ from being a ``standalone'' biosignature as well.  These mechanisms include production via CO$_2$ photolysis as discussed in Sect.~\ref{sec:O2fromO3}, as well as via H$_2$O photolysis either from an extremely active pre-main sequence star \citep{luge15,tian15} or an atmosphere that has allowed H$_2$O to enter the stratosphere due to a lack of cold trap from low abundances of non-condensable gases \citep{word14}.  Ruling out these scenarios could be possible by detections or non-detections of  gaseous species that would be produced or destroyed during these processes.  For example, \om\ and \oz\ abiotic buildup from CO$_2$ photolysis could be revealed via a detection of CO, sometimes called an ``antibiosignature'' (for detailed descriptions of these mechanisms and their spectral discriminants see \citealt{mead17,mead18}).  

Potential abiotic production will require contextual knowledge of an atmosphere to use either \om\ or \oz\ as a biosignature.  Abiotic buildup from CO$_2$ photolysis with high FUV/NUV flux ratio stellar hosts could potentially impact \oz\ more than \om, as it is possible to accumulate \oz\ more easily than \om\ in this scenario, and could potentially allow simultaneous detection of abiotic \oz\ and CH$_4$ under certain conditions \citep{doma14}.  However, predictions of abiotic  \om\ and \oz\ buildup are dependent on the lower boundary conditions of the model in question, so these estimates vary \citep{hu12,doma14,tian14,harm15,gao15}.  Detections or non-detections of CO and CO$_2$ will be important especially in assessing the origin of an \oz\ detection.

The second main argument against \om\ as a reliable biosignature (even when accounting for abiotic sources) is that \om\ levels have only been relatively high for a short period of Earth's geological history.  Oxygenic photosynthesis is thought to have been first used by cyanobacteria $\sim$2.7 Ga , although \om\ buildup during the GOE was not thought to have occurred until $\sim$2.5 Ga (e.g., \citealt{poul20}).  \om\ levels comparable to modern Earth were not reached until the Phanerozoic (541 Ma - present day) sparked by the Cambrian explosion when land began to be colonized by plants \citep{lent17,dahl20}.  Before the GOE \om\ levels were expected to be well below 10$^-3$\% PAL, and potentially remained relatively low during the majority of the Proterozoic (2.5 Ga - 541 Ma) with estimates ranging from $\sim$0.3-10\% PAL \om\ (e.g.,\ \citealt{catl18}).  Even if the lowest \om\ estimates for the Proterozoic were reality ($\sim$0.01\% PAL \om), there is evidence of an \oz\ layer after 2.4 Ga \citep{croc18}.  Although an \om\ detection would be extremely difficult at this abundance, it has been suggested that \oz\ could reveal this undetectable \om\ (e.g., \citealt{lege93,desm02,segu03,lege11,harm15}).  Results of this paper only further prove this point, especially around hotter stars (see Fig.~\ref{fig:O3-O2_MR_all}).  For the G0V and Sun-hosted planets even at a level of 0.01\% PAL \om, they still produce  $\sim$15\% the amount of \oz\ they do at 100\% PAL \om.  This number falls to $\sim$2\% for the M5V-hosted planet, although it still demonstrates a less drastic decrease in \oz\ as \om\ decreases.  This implies that especially for planets orbiting hotter stars that \oz\ is a much longer lived detectable biosignature than \om\ for an Earth-like planet, as detections may be sensitive to Proterozoic \oz\ levels.

Although gaining precise information about \om\ abundance from an \oz\ measurement will be extremely difficult or not possible (see discussion in Sect.~\ref{sec:O2fromO3}), knowing \oz\ abundance alone would still provide valuable information about the atmosphere.  There appears to be a ``bistability limit'' for atmospheric \om, implying that certain \om\ levels would not be stable in the atmosphere due to \om\ sinks and geochemical cycles, as atmosphere switches from reduced to oxidizing (e.g., \citealt{segu03,gold06}). \cite{greg21} calculated that there are only a few stable solutions with \om\ abundances between 3$\times10^{-6}$ and 1\% PAL for an Earth-like planet orbiting the Sun. The existence of this ``bistability limit'' could explain the $\sim$300 Myr delay between the advent of oxygenic photosynthesis and appreciable \om\ accumulation in the atmosphere \citep{gold06}.  \oz\ abundance drastically falls off for all our host stars under 0.01\% PAL \om, implying that an \oz\ detection would allow us to distinguish between pre- and post-GOE \om\ levels with relative ease.  In the search for Earth-like planets, \oz\ appears to be a viable biosignature for a much longer portion of Earth's history, potentially allowing us to infer the existence of oxygenic photosynthesis for much longer than an \om\ detection.

Although \oz\ is not created by life, its UV shielding capabilities could allow estimates of whether the surface environment is safe for life. As seen in Fig.~\ref{fig:surfUV} and Table~\ref{tab:UV_all}, the amount of UV flux reaching the planetary surface begins to converge quickly to the UV incident upon the planet when \om\ abundance drops below 1\% PAL, and it has been predicted that \om\ levels less than this will not be efficient at preventing DNA damage (e.g., \citealt{segu03}).  If an upper \om\ ``bistability limit'' indeed exists at 1\% PAL \citep{greg21}, a detection of \oz\ in a planetary atmosphere could imply a certain amount of UV shielding, and potential surface life.  

However, it is important to remember that the first evidence of life dates back 3.7 Ga, long before the GOE, or even oxygenic photosynthesis \citep{rosi99}.  Although \oz\ may be a longer lived biosignature than \om\ and can indicate substantial UV surface shielding, a non-detection of \oz\ (or \om) could not rule out the existence of life.  Although the surface UV environment would have been harsh before the GOE, it is possible that without a UV screen that life could thrive in the photic zone of the ocean, and perhaps colonize land \citep{cock07}.  It has even been suggested that a significant amount of UV may have been necessary to synthesize prebiotic molecules (e.g., \citealt{pate15,ranj16,rimm18}).  Even substituting \oz\ for \om\ in biosignature searches, life on pre-GOE Earth would be undetectable.

\section{Conclusions}

In this first part of our paper series we show that the nonlinear O$_2$-O$_3$ relationship varies significantly for model atmospheres of planets orbiting different types of host stars, with different trends for planets with hotter host stars versus those with cooler host stars.  As seen in Fig.~\ref{fig:O3-O2_MR_all}, planets orbiting hotter host stars display peak \oz\ abundance at lower \om\ levels than modern Earth, while planets with cooler hosts have \oz\ decrease along with \om.  The increase in \oz\ at lower \om\ levels for hotter host stars is due to the \oz\ layer shifting downward in the atmosphere as \om\ levels (and its ability to absorb UV) decrease.  At these deeper and denser levels of the atmosphere the 3-body reaction that creates \oz\ (Reaction~\ref{r:O2M}) allows more efficient \oz\ production than at high \om\ abundances.  Cooler stars do not experience this effect since it requires a stronger incident FUV flux to push \oz\ formation deep enough in the atmosphere to allow for much faster production.

As \om\ decreases in the atmosphere, photolysis of many gaseous species as well as \od\ production will occur at lower atmospheric levels. The biologically relevant molecules H$_2$O, CH$_4$, and N$_2$O all experience upper atmospheric depletion to different degrees, with these effects significantly more prominent around host stars with higher UV fluxes (Fig.~\ref{fig:genchem}).  As \om\ decreases and photolysis rates increase, HO$_x$ species are primarily pushed to lower altitudes rather than destroyed, while NO$_x$ species are destroyed by photolysis as well as produced at lower atmospheric levels (Fig.~\ref{fig:catalytic}).

UVA and UVB wavelengths are only partially shielded by \oz, so the amount of photons in this wavelength range that reach the planetary surface scales with the amount of incident UVA and UVB flux.  However, for biologically damaging UVC photons, there is a much stronger dependence on both \om\ and \oz\ abundance.  For high \om\ levels, our hottest host star (G0V) had the lowest surface UVC flux, despite also having the highest incident UVC flux (Fig.~\ref{fig:surfUV}).  As \om\ and \oz\ abundances decreases these UVC surface levels begin to converge to incident UVC flux as UV shielding becomes inefficient.

Ozone  features in planetary emission spectra were found to require knowledge of the atmospheric temperature profiles, as the depth of  features is dictated by the temperature difference between the emitting and absorbing layers of the gaseous species.  Since \oz\ NUV absorption is a significant source of stratospheric heating, a large amount of \oz\ along with significant incident NUV flux will cause a smaller temperature difference between the stratosphere and planetary surface, resulting in a shallower spectral feature (Fig.~\ref{fig:emission}). For cooler stars with slower \oz\ production, and therefore less stratospheric temperature inversion, \om\ spectral  features are more intuitive to interpret.  Overall it is clear that interpreting any observation of \oz\ will require the UV spectrum of the host star as well as photochemical and climate modeling of the planetary atmosphere.

Now that we have explored the \om-\oz\ relationship and its impact on planetary emission spectra, let us now return again to our original question: is \oz\ a reliable proxy for \om?  In short, the complicated nature of the \om-\oz\ relationship tells us that \oz\ is not a reliable tracer of \om.  Our results show us that for hotter stars, using \oz\ as a precise tracer for \om\ will not be possible due to the degeneracies in the \om-\oz\ relationship, and for cooler stars it will be very difficult without knowledge of the UV spectrum of the host star as well as planetary atmospheric composition.  However, an \oz\ measurement on its own is still an insightful measurement, even if it does not provide precise information on \om\ abundance. Not only is \oz\ detectable in trace amounts (unlike \om), it additionally allows for an assessment of the UV surface environment of the planet.

There is likely no ``standalone'' atmospheric biosignature, but either \om\ or \oz\ along with atmospheric context could provide evidence of oxygenic photosynthesis.  Both will require detections of other gaseous species to rule out various mechanisms for abiotic \om\ or \oz\ production. Although it is important to note that a CO$_2$ rich planet may be more likely to have an abiotic buildup of \oz\ without \om\ - a strike against \oz\ as a biosignature gas (see Sect.~\ref{sec:o2o3relationship} for detailed discussion).  However, there is a strike against \om\ in comparison to \oz: \om\ has only existed on Earth in relatively high amounts for a short fraction of Earth's geological history.  Ozone, on the other hand, is detectable in trace amounts, potentially making it a longer lived detectable biosignature for Earth-like planets.  It seems that neither \om\ or \oz\ is inherently a ``better'' biosignature than the other; simply that they give different information and can be more or less useful depending on the scenario.  With knowledge of the UV spectrum of the host star along with careful climate and photochemistry modeling we can begin to understand the \om-\oz\ relationship and use \oz\ as a reliable indicator for oxygenic photosynthesis.

\begin{acknowledgements}
We thank both the anonymous referee and journal editor for their comments which helped improve the clarity of this manuscript.  All computing was performed on the HPC cluster at the Technical University of Denmark \citep{hpc}.  This project is funded by VILLUM FONDEN. 
\end{acknowledgements}

\bibliographystyle{aa}
\bibliography{main}{}

\end{document}